\documentclass[twocolumn]{aastex631}

\usepackage{amsmath}
\graphicspath{{../../../../../../../../../../../../Documents/Redshift_Paper/}}
\newcommand{\othree}{[O~{\sc iii}]}
\newcommand{\mgtwo}{Mg~{\sc ii}}
\newcommand{\cfour}{C~{\sc iv}}
\newcommand{\hb}{H$\beta$}
\newcommand{\ha}{H$\alpha$}
\newcommand{\fetwo}{Fe~{\sc ii}}

\newcommand{\hd}{H$\delta$}
\newcommand{\hg}{H$\gamma$}

\newcommand{\otwo}{[O~{\sc ii}]}
\newcommand{\nethree}{[Ne~{\sc iii}]}

\begin{document}


\title{Gemini Near Infrared Spectrograph - Distant Quasar Survey: Augmented Spectroscopic Catalog and a Prescription for Correcting UV-Based Quasar Redshifts}

\email{brandonmatthews@my.unt.edu}

\author[0000-0001-8406-4084]{Brandon M. Matthews}
\affil{Department of Physics, University of North Texas,
Denton, TX 76203, USA}

\author[0000-0003-0192-1840]{Cooper Dix}
\affil{Department of Physics, University of North Texas,
Denton, TX 76203, USA}

\author[0000-0003-4327-1460]{Ohad Shemmer}
\affil{Department of Physics, University of North Texas,
Denton, TX 76203, USA}

\author[0000-0002-1207-0909]{Michael S. Brotherton}
\affil{Department of Physics and Astronomy, University of Wyoming,
Laramie, WY 82071, USA}

\author{Adam D. Myers}
\affil{Department of Physics and Astronomy, University of Wyoming,
Laramie, WY 82071, USA}

\author[0000-0003-1562-5188]{I. Andruchow}
\affil{Facultad de Ciencias Astron\'{o}micas y Geofi\'{i}sicas, Universidad Nacional de La Plata, Paseo del Bosque, B1900FWA La Plata, Argentina}
\affil{Instituto de Astrof\'{i}sica de La Plata, CONICET–UNLP, CCT La Plata, Paseo del Bosque, B1900FWA La Plata, Argentina}

\author[0000-0002-0167-2453]{W. N. Brandt}
\affil{Department of Astronomy and Astrophysics, The Pennsylvania State University, University Park, PA 16802, USA}
\affiliation{Institute for Gravitation and the Cosmos, The Pennsylvania State University, University Park, PA 16802, USA}
\affiliation{Department of Physics, 104 Davey Lab, The Pennsylvania State University, University Park, PA 16802, USA}

\author[0000-0001-6217-8101]{S. C. Gallagher}
\affil{Department of Physics \& Astronomy, University of Western Ontario, 1151 Richmond St, London, ON N6C 1T7, Canada}

\author{Richard Green}
\affil{Steward Observatory, University of Arizona, 933 N Cherry Ave, Tucson, AZ 85721, USA}

\author{Paulina Lira}
\affil{Departamento de Astronom\'ia, Universidad de Chile, Casilla 36D, Santiago, Chile}

\author[0000-0003-1081-2929]{Jacob N. McLane}
\affil{Department of Physics and Astronomy, University of Wyoming,
Laramie, WY 82071, USA}

\author[0000-0002-7092-0326]{Richard M. Plotkin}
\affil{Department of Physics, University of Nevada, Reno, NV 89557, USA}
\affil{Nevada Center for Astrophysics, University of Nevada, Las Vegas, NV 89154, USA}

\author[0000-0002-1061-1804]{Gordon T. Richards}
\affil{Department of Physics, Drexel University, 32 S. 32nd Street, Philadelphia, PA 19104, USA}

\author[0000-0001-8557-2822]{Jessie C. Runnoe}
\affil{Department of Physics \& Astronomy, Vanderbilt University, 6301 Stevenson Center Ln, Nashville, TN 37235, USA}

\author[0000-0001-7240-7449]{Donald P. Schneider}
\affil{Department of Astronomy and Astrophysics, The Pennsylvania State University, University Park, PA 16802, USA}
\affiliation{Institute for Gravitation and the Cosmos, The Pennsylvania State University, University Park, PA 16802, USA}

\author[0000-0002-0106-7755]{Michael A. Strauss}
\affil{Department of Astrophysical Sciences, Princeton University, Princeton, NJ 08544, USA}

\revised{2023 March 19}
\received{2022 December 1}
\accepted{2023 April 19}

\begin{abstract}

Quasars at $z \gtrsim 1$ most often have redshifts measured from rest-frame ultraviolet emission lines. One of the most common such lines, \cfour\ $\lambda1549$, shows blueshifts up to $\approx 5000\ \rm{km\  s^{-1}}$, and in rare cases even higher. This blueshifting results in highly uncertain redshifts when compared to redshift determinations from rest-frame optical emission lines, e.g., from the narrow \othree\ $\lambda5007$ feature. We present spectroscopic measurements for 260 sources at \hbox{$1.55 \lesssim z \lesssim 3.50$} having \hbox{$-28.0 \lesssim M_i \lesssim -30.0$ mag} from the Gemini Near Infrared Spectrograph - Distant Quasar Survey (GNIRS-DQS) catalog, augmenting the previous iteration which contained 226 of the 260 sources whose measurements are improved upon in this work. We obtain reliable systemic redshifts based on \othree\ $\lambda5007$ for a subset of 121 sources which we use to calibrate prescriptions for correcting UV-based redshifts. These prescriptions are based on a regression analysis involving \cfour ~full-width-at-half-maximum intensity and equivalent width, along with the UV continuum luminosity at a rest-frame wavelength of 1350 \AA. Applying these corrections can improve the accuracy and the precision in the \cfour -based redshift by up to $\sim850~\rm{km\ s^{-1}}$ and $\sim150~\rm{km\ s^{-1}}$, respectively, which correspond to $\sim8.5$ Mpc and \hbox{$\sim1.5$ Mpc} in comoving distance at $z=2.5$. Our prescriptions also improve the accuracy of the best available multi-feature redshift determination algorithm by $\sim100~\rm{km\ s^{-1}}$, indicating that the spectroscopic properties of the \cfour\ emission line can provide robust redshift estimates for high-redshift quasars. We discuss the prospects of our prescriptions for cosmological and quasar studies utilizing upcoming large spectroscopic surveys.

\end{abstract}


\section{Introduction} \label{sec:intro}

Obtaining systemic redshifts ($z_{\rm sys}$) for quasars to accuracies better than $1000~\rm{km\ s^{-1}}$ is necessary for a variety of reasons. These include measuring the kinematics of outflowing material near the supermassive black hole (SMBH) that impact star formation rates in the quasar's host galaxy \citep[e.g.,][]{2010MNRAS.401....7H,2012MNRAS.425L..66M,2018agn..confE..68C}, and cosmological studies that utilize redshifts as distance indicators, such as quasar clustering and the proximity effect at high redshift \citep[e.g.,][]{1979Natur.281..358A,1999astro.ph..5116H,2007AJ....133.2222S,2013AJ....145...10D,2016AJ....151...61M,2019MNRAS.482.3497Z}.

A quasar $z_{\rm sys}$ value is typically determined from spectroscopy in the optical band relying, particularly, on the wavelength of the peak of the narrow \hbox{\othree\ $\lambda5007$} emission line at $z \lesssim 0.8$, the \hbox{\mgtwo\ $\lambda\lambda2798, 2803$} doublet for $0.4 \lesssim z \lesssim 2.3$, or the Balmer lines up to $z \sim 1$, in order of increasing uncertainty on the derived $z_{\rm sys}$ value, ranging from \hbox{$\sim 50~\rm{km\  s^{-1}}$} to \hbox{$\sim 600~\rm{km\  s^{-1}}$} \citep[e.g.,][]{2005AJ....130..381B,2016ApJ...831....7S,2020ApJ...895...74N}. However, at higher redshifts, these $z_{\rm sys}$ indicators shift out of the optical band, and redshift determinations usually rely on shorter wavelength, and typically higher ionization emission lines such as \cfour ~$\lambda1549$. Such emission lines are known to show additional kinematic offsets of up to several 10$^3$ $\rm{km\  s^{-1}}$ that add uncertainties of this magnitude to the derived redshift values \citep[e.g.,][]{1982ApJ...263...79G,1992ApJS...79....1T,2009ApJ...692..758G,2016ApJ...831....7S,2018A&A...617A..81V}. The redshifts of distant quasars determined from large spectroscopic surveys \citep[e.g., Sloan Digital Sky Survey, SDSS,][]{2000AJ....120.1579Y,2016SPIE.9908E..1MT,2016arXiv161100036D,2020ApJS..250....8L}, that are limited to $\lambda_{\rm obs} \lesssim1 ~\mu$m, therefore will have uncertainties on the order of tens of Mpc at $z=2.5$, when converting from velocity space into comoving distance \citep[e.g.,][]{2013JCAP...05..018F}.

A direct comparison of SDSS Pipeline redshifts \citep{2012AJ....144..144B,2020ApJS..250....8L} with $z_{\rm sys}$ values obtained from rest-frame optical indicators show that corrections to UV-based redshifts can be made despite the presence of potentially large uncertainties. Past investigations such as Hewett \& Wild (\citeyear{2010MNRAS.405.2302H}), Mason et al. (\citeyear{2017MNRAS.469.4675M}) and Dix et al. (\citeyear{2020ApJ...893...14D}), hereafter HW10, M17, and D20, respectively, have demonstrated that these uncertainties can be mitigated through corrections obtained from regression analyses based on pre-existing rest-frame UV-optical spectral properties and used as prescriptions for correcting UV-based redshifts.

HW10 relied primarily on sampling methods wherein an average quasar spectrum was generated using a large sample of existing quasar spectra, and then statistical analysis was used to provide offsets for any given quasar with respect to this ``master" spectrum in order to correct for any uncertainties. However, this offset correction becomes less reliable for high redshift quasars as important emission lines such as \othree ~and \mgtwo ~leave the optical band, and so additional corrections are needed \citep[see, e.g.,][]{2020MNRAS.492.4553R}.

M17 and D20 used regression analyses that apply empirical corrections to UV-based redshifts involving the \cfour\ spectroscopic parameter space,  a diagnostic of quasar accretion power \citep{2011AJ....141..167R,2020MNRAS.492.4553R,2020ApJ...899...96R}, which affects the wavelengths of emission-line peaks. Specifically, these parameters include the rest-frame equivalent width (EW) and full width at half maximum intensity (FWHM) of the \cfour\ line\footnote{We discuss additional velocity width measurement methods in Appendix~\ref{sec:fwhm}.} as well as the continuum luminosity at the base of this line. Such corrections have been applied to sources that lack broad absorption lines and are not radio-loud\footnote{\label{note:radio}We consider radio-loud quasars to have $R >$ 100, where $R$ is defined as $R=f_{\nu}$(5~GHz) / $f_{\nu}$ (4400~\AA), where $f_{\nu}$(5~GHz) and $f_{\nu}$(4400~\AA) are the flux densities at a rest-frame frequency of 5~GHz and a rest-frame wavelength of 4400~\AA, respectively \citep{1989AJ.....98.1195K}} in order to minimize the effects of absorption and continuum boosting, respectively, to the \cfour\ line profile to mitigate potential complications arising from these sources and provide the most reliable results possible.

The D20 analysis, an extension of the M17 study, was based on a non-uniform sample of 55 SDSS sources with spectral coverage in the rest-frame optical and UV. Here, we use a larger and more uniform sample of 121 sources with highly reliable $z_{\rm sys}$ values drawn from an augmentation of the Gemini Near Infrared Spectrograph - Distant Quasar Survey (GNIRS-DQS) near-infrared (NIR) spectral inventory \citep[][hereafter M21]{2021ApJS..252...15M}. Our results allow us to obtain significantly improved prescriptions for correcting UV-based redshifts. Section~\ref{sec:sample} describes the properties of the quasar sample and the respective redshift measurements, along with an augmentation of the M21 catalog of spectral properties from GNIRS-DQS. Section~\ref{sec:data} presents prescriptions for correcting UV-based quasar redshifts based on multiple regression analyses including several velocity width indicators, alongside discussion of the redshift dependence of the velocity offset corrections, and redshift estimates for quasars with extremely high velocity offsets. Our conclusions are presented in Section~\ref{sec:app}. Throughout this paper we adopt a flat $\Lambda$CDM cosmology with $\Omega_\Lambda = 1 - \Omega_{\rm{M}} = 0.7$ and $H_0 = 70$ ${\rm km\ s^{-1}} ~{\rm Mpc^{-1}}$ \citep[e.g.,][]{2007ApJS..170..377S}.

\section{Sample Selection} \label{sec:sample}

\begin{figure*}[]
\includegraphics[scale=0.67]{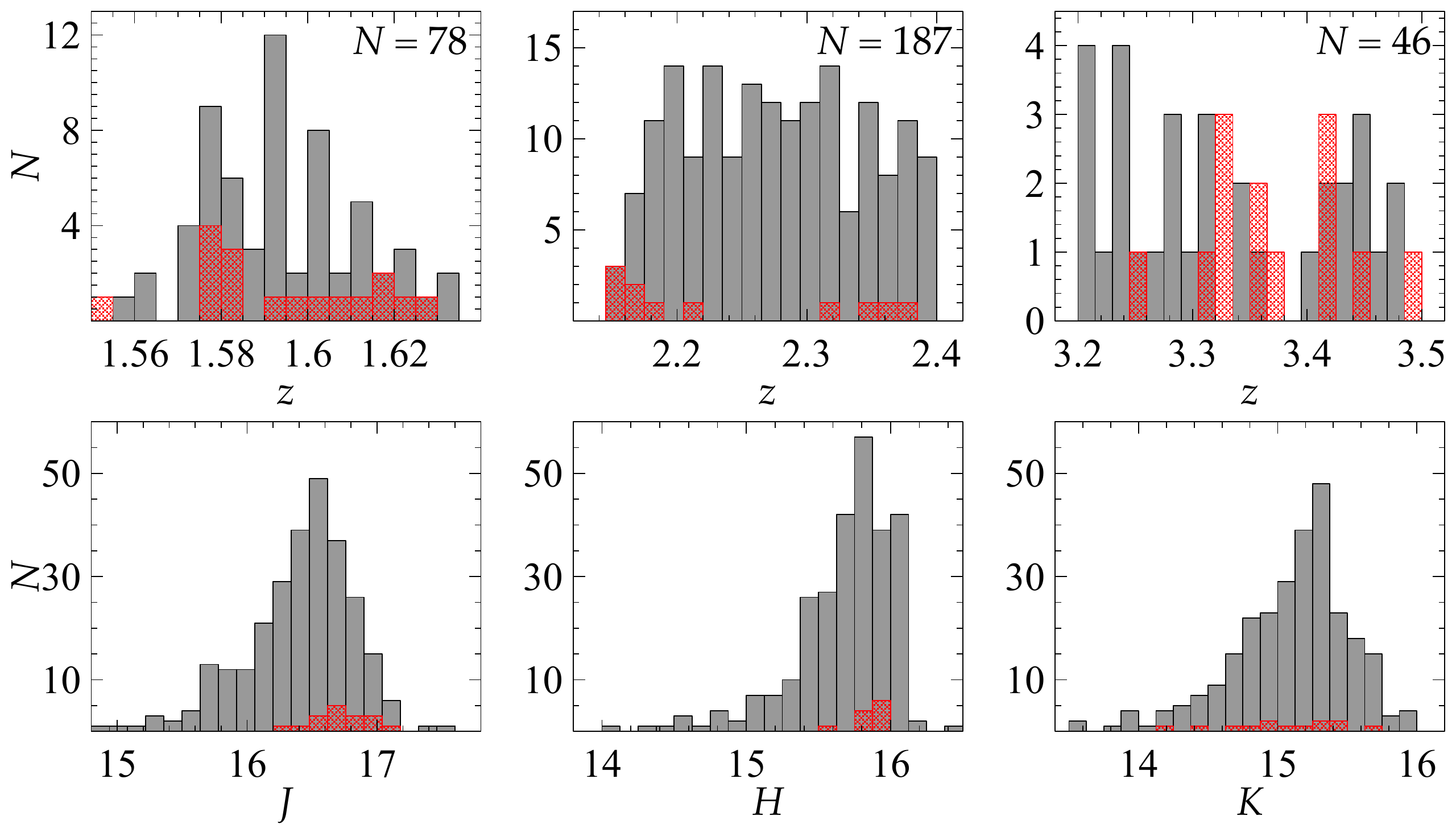}\caption{Distributions of the redshift estimate from SDSS \citep[][Table D1, column 27 ``Z"]{2020ApJS..250....8L} in each redshift interval (top), and corresponding magnitude distributions (bottom). The initial GNIRS-DQS sample is marked in  grey, and sources from the augmented sample are shown in red. The three redshift bins correspond to the \hb ~and \othree ~lines appearing at the center of the $J$, $H$, or $K$ photometric bands. The number of sources observed in each redshift bin is marked in each of the top panels. Of a total of 311 sources observed, 272 of which were reported in M21, usable NIR spectra were obtained for 260 sources ($\sim84\%$ completion rate); the NIR spectra of 226 of these were presented in M21 and the remaining 34 are presented in this work.}
\label{fig:jhk}
\end{figure*}

\begin{figure}[]
\includegraphics[scale=0.5]{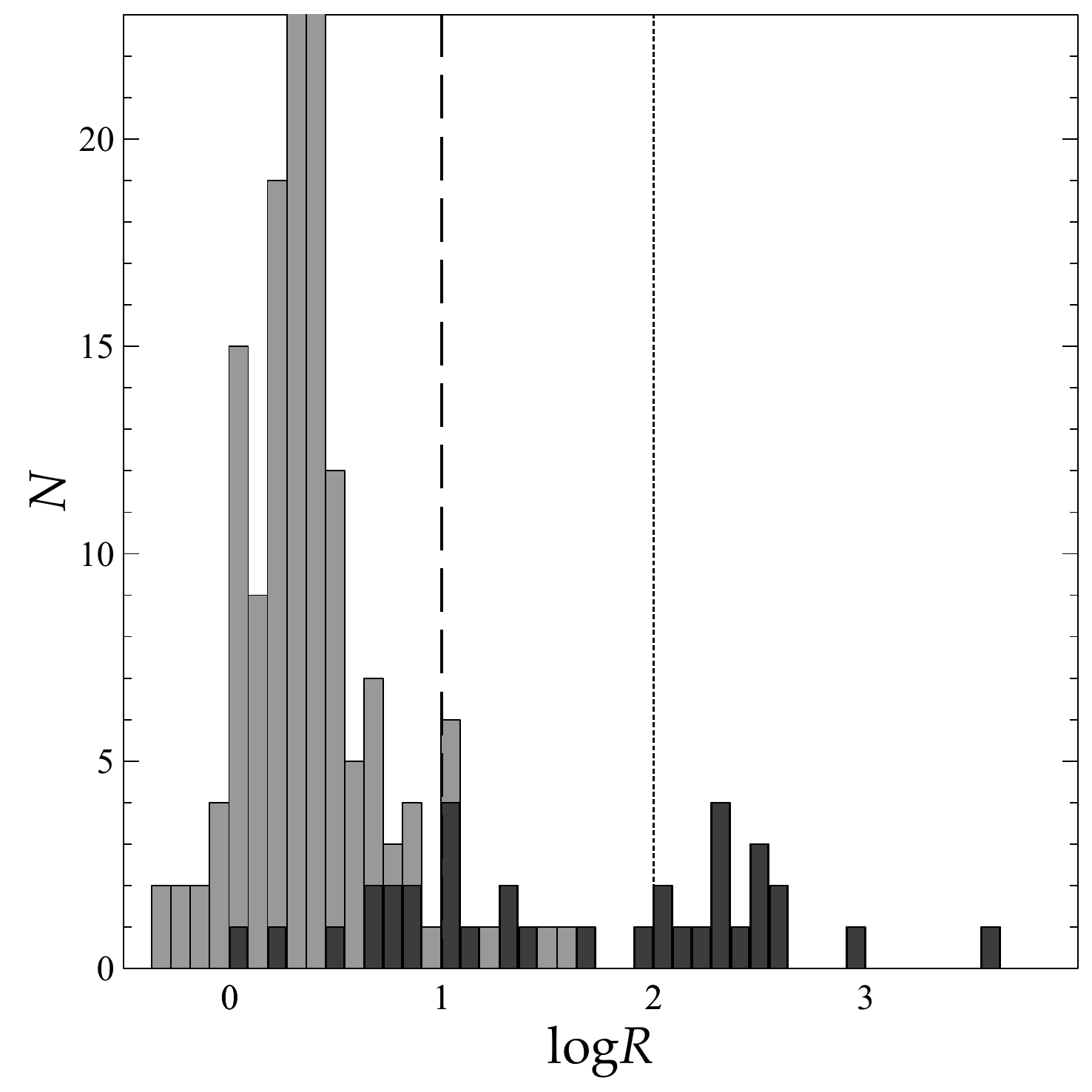}\caption{Radio-loudness distribution of the GNIRS-DQS sources. Darker shaded regions indicate new sources not in M21. The dashed line at \hbox{log \textit{R} = 1} indicates the threshold for radio-quiet quasars, and the dotted line at \hbox{log \textit{R} = 2} indicates the threshold for radio-loud quasars (see also M21).}
\label{fig:RL}
\end{figure}

\begin{figure}[]
\includegraphics[scale=0.5]{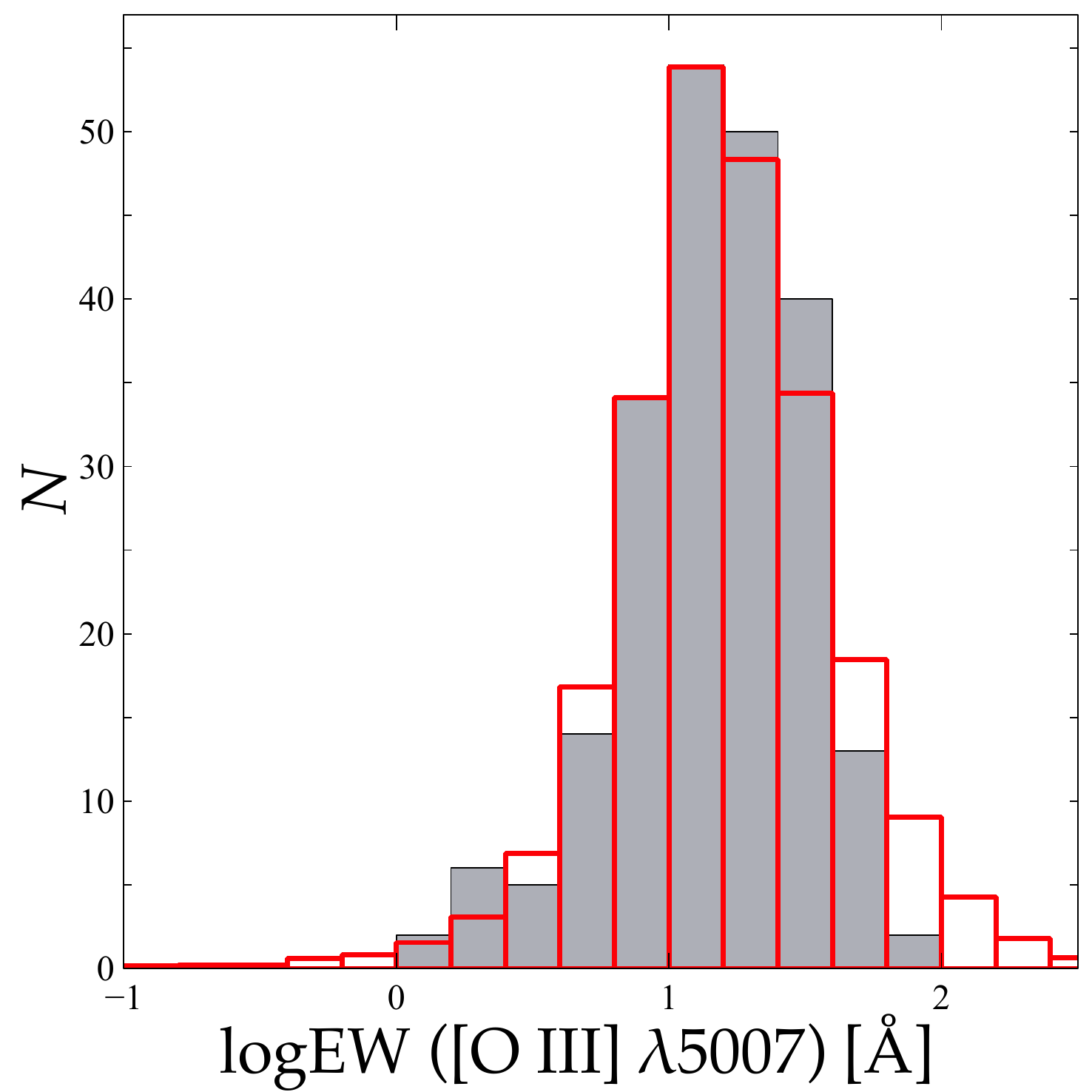}\caption{\othree\ $\lambda5007$ rest-frame EW distribution of 220 GNIRS-DQS sources (solid gray histogram) and a similar distribution from Shen et al. (\citeyear{2011ApJS..194...45S}) (red outline; scaled down by a factor of 500). See M21 for additional discussion. Forty sources did not meet our threshold of reliability for an \othree\ EW measurement that we require to be greater than \hbox{$1$~\AA}.}
\label{fig:o3ew}
\end{figure}

Our quasar sample is drawn from GNIRS-DQS, which comprises the largest, most uniform sample of optically selected high-redshift quasars having NIR spectroscopic coverage (M21). The GNIRS-DQS sources were selected from all SDSS quasars \citep{2018A&A...613A..51P,2020ApJS..250....8L} having \hbox{$-28.0 \lesssim M_i \lesssim -30.0$ mag} at \hbox{$1.55 \lesssim z \lesssim 3.50$} for which the \hb\ and \othree\ emission-lines can be covered in either the $J$, $H$, or $K$ bands, spanning a monochromatic luminosity ($\lambda L_{\lambda}$) at 5100 \AA\ in the range of \hbox{$\sim10^{46} - 10^{47}~\rm{erg} \rm{s^{-1}}$ $\rm \AA^{-1}$}.  We augment the original GNIRS-DQS sample with 34 additional sources, selected in a similar fashion as described below, and shown in Figure~\ref{fig:jhk}. Distributions of radio loudness and \othree~$\lambda5007$ EW for the GNIRS-DQS sources are shown in Figures~\ref{fig:RL} and~\ref{fig:o3ew}, respectively.

\subsection{The Augmented GNIRS-DQS Catalog}

We present spectroscopic observations for 34 sources that were observed in semester 2020B as part of our GNIRS-DQS campaign (see M21 for a detailed description of the observational strategy and the instrument configuration). We also include spectroscopic observations for 11 sources that were observed in a similar fashion, albeit with a narrower slit, 0.30$''$, in semester 2015A (program GN-2015A-Q-68; PI: Brotherton). Three of the 2020B sources, and one of the 2015A sources, were observed twice in their respective semesters. The log of these additional observations is given in Table~\ref{tab:obslog}.

Three of the 2020B observations were repeat observations of SDSS J094427.27+614424.6, SDSS J113924.64+332436.9, and SDSS J223934.45-004707.2, that were previously described in M21; the repeat observations were intended to improve the quality of the spectroscopic data for these three sources (see Table~\ref{tab:obslog}). Four of the 2020B observations, and 4 of the 2015A observations did not yielded adequate spectroscopic data, and are marked appropriately in Table~\ref{tab:obslog}. Therefore, in total, we have 37 useful observations, three of which are of sources that were already presented in M21. The augmented GNIRS-DQS catalog comprises of a total of 260 sources, 34 of which have been added in this work.

The formatting for the basic spectral properties of all 260 GNIRS-DQS objects is presented in Tables~\ref{tab:obj} and~\ref{tab:supp} in a similar fashion to Tables 2 and 3 in M21. The GNIRS-DQS sample was originally selected from the SDSS quasar catalogs for Data Release (DR) 12 and DR14 \citep{2017A&A...597A..79P,2018A&A...613A..51P}; the augmented GNIRS-DQS catalog presented here includes 27 sources that were selected from SDSS DR16 \citep{2020ApJS..250....8L}, which are marked appropriately in Table~\ref{tab:obslog}, and 7 sources from GN-2015A-Q-68 that were selected from SDSS DR7 \citep{2010AJ....139.2360S}. DR16 measurements have been adopted for the full sample \citep{2020ApJS..250....8L}. Table~\ref{tab:gauss} presents the parameters used to model all of the emission lines, using Gaussian profiles, in the GNIRS-DQS spectra. For each profile, these parameters include the observed-frame wavelength of the line peak, velocity width (FWHM), and flux-density normalization ($f_{\lambda}$). All of the GNIRS spectra and their best-fit models are available electronically at\dataset[NOIRlab]{https://datalab.noirlab.edu/gnirs_dqs.php}\footnote{https://datalab.noirlab.edu/gnirs\_dqs.php}.

\begin{deluxetable*}{lccccccccc}
\tablecolumns{10}
\tabletypesize{\scriptsize}
\tablewidth{0pc}
\tablecaption{GNIRS-DQS Observation Log \label{tab:obslog}}
\tablehead
{
\colhead{Quasar} &
\colhead{$z_{\rm SDSS}$\tablenotemark{a}} &
\colhead{$J$} &
\colhead{$H$} &
\colhead{$K$} &
\colhead{Obs. Date} &
\colhead{Net Exp.} &
\colhead{Comments} &
\colhead{BAL} &
\colhead{RL} \\
\colhead{} &
\colhead{} &
\colhead{[mag]} &
\colhead{[mag]} &
\colhead{[mag]} &
\colhead{} &
\colhead{[s]} &
\colhead{} &
\colhead{} &
\colhead{} \\
\colhead{(1)} &
\colhead{(2)} &
\colhead{(3)} &
\colhead{(4)} &
\colhead{(5)} &
\colhead{(6)} &
\colhead{(7)} &
\colhead{(8)} &
\colhead{(9)} &
\colhead{(10)}
}
\startdata
SDSS J001018.88+280932.5*  & 1.612    & 16.56   & 15.80   & 15.76   & 2020 Dec 09                & 1800    &\nodata    & \nodata & \nodata            \\
SDSS J003001.11$-$015743.5  & 1.582    & 17.08   & 15.96   & 15.76   & 2020 Sep 09                & 1800    &\nodata    & \nodata & \nodata            \\
SDSS J003853.15+333044.3  & 2.357    & 16.81   & 15.98   & 15.29   & 2020 Dec 25                & 1800    &\nodata   & \nodata & \nodata            \\
SDSS J004613.54+010425.7  & 2.150    & 16.44   & 15.85   & 15.02   & 2020 Dec 11                & 1800    &\nodata    & 1 & \nodata            \\
SDSS J004710.48+163106.5  & 2.165    & 16.33   & 15.62   & 14.90   & 2020 Dec 11                & 1800    &\nodata    & \nodata & \nodata            \\
SDSS J005307.71+191022.7*  & 1.583    & 16.72   & 15.79   & 15.43   & 2020 Sep 08                & 1800    &\nodata    & \nodata & \nodata            \\
SDSS J020329.86-091020.3*  & 1.579    & 17.02   & 15.97   & 15.64   & 2020 Aug 23                & 900    &\nodata    & \nodata & \nodata            \\
&\nodata & \nodata      & \nodata  & \nodata  & 2020 Sep 11 & 900       &\nodata     & \nodata & \nodata     \\
SDSS J073132.18+461347.0*  & 1.578    & 16.71   & 15.83   & 15.31   & 2020 Sep 29                & 1350    &\nodata   & \nodata & \nodata            \\
SDSS J080117.91+333411.9*  & 1.598    & 16.73   & 15.99   & 15.79   & 2020 Oct 05                & 1350    &\nodata     & \nodata & \nodata            \\
SDSS J080429.61+113013.9*  & 2.165    & 16.64   & 15.99   & 15.13   & 2020 Nov 27                & 1800    & 2     & \nodata & \nodata            \\
SDSS J080636.81+345048.5*  & 1.553    & 16.45   & 15.88   & 16.58   & 2020 Sep 30                & 1800    &\nodata      & 1 & \nodata            \\
SDSS J080707.37+260729.1*  & 2.312    & 16.84   & 15.99   & 15.53   & 2020 Sep 30                & 1800    & 2     & \nodata & \nodata            \\
SDSS J081520.94+323512.9*  & 1.584    & 16.90   & 15.85   & 15.55   & 2020 Nov 28                & 1800    &2      & \nodata & \nodata            \\
SDSS J084017.87+103428.8 &  3.330   & 16.69   & 16.47   & 15.27    & 2015 Apr 23           & 1720    &\nodata    & \nodata & \nodata            \\
SDSS J084401.95+050357.9 &  3.350  & 15.39   & 14.93   & 14.19    & 2015 Apr 06           & 800    &\nodata    & \nodata & \nodata            \\
SDSS J084526.75+550546.8*  & 1.620    & 16.33   & 15.65   & 15.18   & 2020 Nov 27               & 1800    &\nodata      & \nodata & \nodata            \\
SDSS J091425.72+504854.9*  & 2.341    & 17.18   & 15.98   & 15.17   & 2020 Nov 29               & 1800    &\nodata      & \nodata & \nodata            \\
SDSS J092942.97+064604.1*  & 1.608    & 16.65   & 15.53   & 15.28   & 2020 Nov 30               & 1800    & \nodata     & \nodata & \nodata            \\
SDSS J094140.16+325703.2*  & 3.452    & 16.55   & 15.81    & 15.24   & 2020 Nov 29               & 1800    &\nodata     & \nodata & \nodata            \\
SDSS J094427.27+614424.6*  & 2.340    & 16.41   & 15.61   & 14.72   & 2020 Dec 09                & 1800    & 3     & \nodata & \nodata            \\
SDSS J095047.45+194446.1*  & 1.575    & 16.80   & 15.98   & 15.62   & 2020 Dec 12                & 900    &\nodata      & \nodata & \nodata            \\
&\nodata & \nodata      & \nodata  & \nodata  & 2020 Dec 21 & 900       &\nodata     & \nodata & \nodata     \\
SDSS J095555.68+351652.6*  & 1.616    & 16.99   & 15.97   & 15.85   & 2020 Dec 09                & 1800    &\nodata      & \nodata & \nodata            \\
SDSS J101724.26+333403.3*  & 1.579    & 16.49   & 15.84   & 15.40   & 2020 Nov 30                & 1800    &\nodata      & \nodata & \nodata            \\
SDSS J111127.43+293319.3*  & 2.178    & 16.42   & 15.88   & 15.10   & 2020 Dec 31                & 1800    &2      & \nodata & \nodata            \\
SDSS J112726.81+601020.2*  & 2.159    & 16.60   & 15.79   & 15.40   & 2020 Dec 31                & 2250   & \nodata      & \nodata & \nodata            \\
SDSS J112938.46+440325.0*  & 2.213    & 16.99   & 15.88   & 15.11   & 2021 Jan 02                & 1800    &\nodata      & 1 & \nodata            \\
SDSS J113330.17+144758.8*  & 3.248    & 16.90   & 15.88   & 15.64   & 2021 Jan 02                & 1800    &\nodata      & 1 & \nodata            \\
SDSS J113924.64+332436.9*  & 2.314    & 16.38   & 15.95   & 14.85   & 2020 Dec 09               & 1800    & 3   & \nodata & \nodata            \\
SDSS J122343.15+503753.4 & 3.491  & 15.90   & 15.57   & 14.69    & 2015 Mar 30           & 1160    &\nodata    & \nodata & \nodata            \\
SDSS J122938.61+462430.5*  & 2.152    & 16.30   & 15.77   & 15.19   & 2020 Nov 30                & 1800    &\nodata      & \nodata & \nodata            \\
SDSS J130213.54+084208.6 & 3.305  & 16.12   & 15.64    & 15.02    & 2015 Apr 01           & 1720    &2  & \nodata & \nodata            \\
SDSS J131048.17+361557.7 & 3.420   & 15.79   & 15.11   & 14.38    & 2015 Apr 05           & 800    & 2    & \nodata & \nodata            \\
SDSS J132845.00+510225.8 & 3.411  & 16.10   & 15.53    & 14.77   & 2015 Apr 05           & 1160    &\nodata    & \nodata & \nodata            \\
SDSS J141321.05+092204.8 & 3.327  & 16.16   & 15.63   & 15.05   & 2015 Apr 05           & 1160    &\nodata    & \nodata & \nodata            \\
SDSS J142123.97+463318.0 & 3.378  & 16.28    & 15.49    & 14.89   & 2015 Apr 07           & 1700    & 2   & \nodata & \nodata            \\
SDSS J142755.85$-$002951.1 & 3.362   & 16.60   & 15.91   & 15.27   & 2015 Apr 01           & 1720    &\nodata    & \nodata & \nodata            \\
SDSS J165523.09+184708.4 & 3.327   & 16.28    & 15.88    & 15.19   & 2015 Apr 08           & 1720    & 2    & \nodata & \nodata            \\
SDSS J173352.23+540030.4 & 3.424  & 15.87   & 15.72    & 14.95   & 2015 Mar 23           & 1190    &\nodata    & \nodata & \nodata            \\
&\nodata & \nodata      & \nodata  & \nodata  & 2015 Apr 01 & 680       &\nodata     & \nodata & \nodata     \\
SDSS J210558.29$-$011127.5 & 1.625    & 16.61   & 15.49   & 15.54   & 2020 Aug 21                & 2250 &1      & \nodata & \nodata            \\
SDSS J211251.06+000808.3* & 1.618    & 16.85   & 15.89   & 15.89   & 2020 Aug 19                & 1800 &1      & \nodata & \nodata            \\
SDSS J213655.35$-$080910.1 & 1.591    & 16.96   & 15.56   & 15.74   & 2020 Aug 23                & 1800 &\nodata      & \nodata & \nodata            \\
SDSS J220139.99+114140.8* & 2.382    & 16.87   & 15.76   & 15.84   & 2020 Aug 30                & 1800 &1      & \nodata & \nodata            \\
SDSS J222310.76+180308.1* & 1.602    & 16.70   & 15.99   & 15.60   & 2020 Sep 01                 & 1800 &1      & \nodata & \nodata            \\
SDSS J223934.45$-$004707.2  & 2.121    & 16.91   & 15.97   & 15.70   & 2020 Oct 03              & 1800    & 1, 3      & \nodata & \nodata            \\
SDSS J233304.61$-$092710.9  & 2.121    & 16.17   & 15.41   & 14.83   & 2021 Jan 01              & 1800    &1      & \nodata & \nodata            \\
&\nodata & \nodata      & \nodata  & \nodata  & 2021 Jan 02 & 900       &\nodata     & \nodata & \nodata     \\
\enddata
\tablenotetext{a}{Value based on best available measurement in SDSS DR16 \citep[][Table D1, column 27 ``Z"]{2020ApJS..250....8L}}
\tablenotetext{*}{Denotes object selected from Data Release 16.}
\tablecomments{Several sources have more than one observation, indicated by an empty source name. Only the 2020B and 2015A observations are shown. All 333 observations of GNIRS-DQS are available in the electronic version. All SDSS data taken from DR16.\\
\\
Comments in Column (8) represent:\\
$[1]$ At least one exposure did not meet our observation conditions requirements.\\
$[2]$ Observation failed to provide spectrum of the source due to bad weather, instrument artifacts, or other technical difficulties during the observation.\\
$[3]$ Re-observed and updated from M21.}
\end{deluxetable*}

\subsection{Improved Spectroscopic Inventory}

Tables~\ref{tab:obj} and~\ref{tab:supp} include improved measurements of all spectral features. In particular, they include measurements of the rest-frame optical \fetwo\ emission blend which was fitted for each source in the same manner as in M21; however, each such feature now has a measured EW value and errors, thus effectively removing all the upper limits on the EWs (cf. Table 2 of M21). We fit two Gaussians to each broad emission-line profile to accommodate a possible asymmetry arising from, e.g., absorption, or outflows. We note that the two Gaussian fit per broad emission line is adopted only to characterize the line shape; the two Gaussians do not imply two physically distinct regions. The errors on the spectral measurements were calculated in the same manner as the other uncertainties described in M21, with upper and lower values being derived from a distribution of values recorded during the iterative process of broadening the \fetwo ~template (see M21 for a detailed description of the \fetwo\ blend fitting process).

In addition to the inclusion of 34 new sources, Tables~\ref{tab:obj} and~\ref{tab:supp} contain the most reliable data for the entire GNIRS-DQS sample following re-measurement of each spectrum with additional vetting and visual inspection, particularly with respect to the \othree\ and \fetwo\ fitting. These data therefore supersede the corresponding data presented in M21.

\begin{deluxetable*}{llllll}
\tablecolumns{6}
\tabletypesize{\scriptsize}
\tablewidth{0pc}
\tablecaption{Column Headings for Spectral Measurements \label{tab:obj}}
\tablehead
{
\colhead{Column} &
\colhead{Name} &
\colhead{Bytes} &
\colhead{Format} &
\colhead{Units} &
\colhead{Description}\\
\colhead{(1)} &
\colhead{(2)} &
\colhead{(3)} &
\colhead{(4)} &
\colhead{(5)} &
\colhead{(6)}
}
\startdata
1 & OBJ & (1-24) & A24 & \nodata & SDSS object designation\\
2 & ZSYS & (26-30) & F5.3 & \nodata & Systemic redshifts\\
3 & ZSRC & (32-34) & A3 & \nodata & Emission line source for systemic redshift as described in M21\\
3 & LC\_MG II & (36-38) & I5 & \AA & \mgtwo\ observed-frame wavelength\tablenotemark{a} \\
4 & LC\_MG II\_UPP & (40-41) & I2 & \AA & Upper uncertainty for the line peak of \mgtwo\\
5 & LC\_MG II\_LOW & (43-44) & I2 & \AA & Lower uncertainty for the line peak of \mgtwo\\
6 & FWHM\_MG II & (46-49) & I4 & $\rm{km}$ $\rm{s^{-1}}$  & FWHM of \mgtwo\\
7 & FWHM\_MG II\_UPP & (51-54) & I4 & $\rm{km}$ $\rm{s^{-1}}$  & Upper uncertainty of FWHM of \mgtwo\\
8 & FWHM\_MG II\_LOW & (56-59) & I4 & $\rm{km}$ $\rm{s^{-1}}$  & Lower uncertainty of FWHM of \mgtwo\\
9 & EW\_MG II & (61-62) & I2 & \AA & Rest-frame EW of \mgtwo\\
10 & EW\_MG II\_UPP & (64-65) & I2 & \AA & Upper uncertainty of EW of \mgtwo\\
11 & EW\_MG II\_LOW & (67-68) & I2 & \AA & Lower uncertainty of EW of \mgtwo\\
12 & AS\_MG II & (70-78) & E9.2 & \nodata & Asymmetry of the double Gaussian fit profile of \mgtwo\\
13 & KURT\_MG II & (80-83) & F4.2 & \nodata & Kurtosis of the double Gaussian fit profile of \mgtwo\\
14 & LC\_HB & (85-89) & I5 & \AA & \hb\ observed-frame wavelength\tablenotemark{a} \\
15 & LC\_HB\_UPP & (91-92) & I2 & \AA & Upper uncertainty for the line peak of \hb\\
16 & LC\_HB\_LOW & (94-95) & I2 & \AA & Lower uncertainty for the line peak of \hb\\
17 & FWHM\_HB & (97-101) & I5 & $\rm{km}$ $\rm{s^{-1}}$  & FWHM of \hb\\
18 & FWHM\_HB\_UPP & (103-107) & I5 & $\rm{km}$ $\rm{s^{-1}}$  & Upper uncertainty of FWHM of \hb\\
19 & FWHM\_HB\_LOW & (109-112) & I5 & $\rm{km}$ $\rm{s^{-1}}$  & Lower uncertainty of FWHM of \hb\\
20 & EW\_HB & (114-116) & I3 & \AA & Rest-frame EW of \hb\\
21 & EW\_HB\_UPP & (118-119) & I2 & \AA & Upper uncertainty of EW of \hb\\
22 & EW\_HB\_LOW & (121-122) & I2 & \AA & Lower uncertainty of EW of \hb\\
23 & AS\_HB & (124-132) & E9.2 & \nodata & Asymmetry of the double Gaussian fit profile of \hb\\
24 & KURT\_HB & (134-137) & F4.2 & \nodata & Kurtosis of the double Gaussian fit profile of \hb\\
25 & LC\_O III & (139-143) & I5 & \AA & \othree ~$\lambda$5007 observed-frame wavelength\tablenotemark{a} \\
26 & LC\_O III\_UPP & (145-146) & I2 & \AA & Upper uncertainty for the line peak of \othree ~$\lambda$5007\\
27 & LC\_O III\_LOW & (148-149) & I2 & \AA & Lower uncertainty for the line peak of \othree ~$\lambda$5007\\
28 & FWHM\_O III & (151-154) & I4 & $\rm{km}$ $\rm{s^{-1}}$  & FWHM of \othree ~$\lambda$5007\\
29 & FWHM\_O III\_UPP & (156-159) & I4 & $\rm{km}$ $\rm{s^{-1}}$  & Upper uncertainty of FWHM of \othree ~$\lambda$5007\\
30 & FWHM\_O III\_LOW & (161-164) & I4 & $\rm{km}$ $\rm{s^{-1}}$  & Lower uncertainty of FWHM of \othree ~$\lambda$5007\\
31 & EW\_O III & (166-173) & E8.2 & \AA & Rest-frame EW of \othree ~$\lambda$5007\\
32 & EW\_O III\_UPP & (175-182) & E8.2 & \AA & Upper uncertainty of EW of \othree ~$\lambda$5007\\
33 & EW\_O III\_LOW & (184-191) & E8.2 & \AA & Lower uncertainty of EW of \othree ~$\lambda$5007\\
34 & AS\_O III & (193-201) & E9.2 & \nodata & Asymmetry of the double Gaussian fit profile of \othree ~$\lambda$5007\\
35 & KURT\_O III & (203-206) & F4.2 & \nodata & Kurtosis of the double Gaussian fit profile of \othree ~$\lambda$5007\\
36 & LC\_HA & (208-212) & I5 & \AA & \ha\ observed-frame wavelength\tablenotemark{a} \\
37 & LC\_HA\_UPP & (214-215) & I2 & \AA & Upper uncertainty for the line peak of \ha\\
38 & LC\_HA\_LOW & (217-218) & I2 & \AA & Lower uncertainty for the line peak of \ha\\
39 & FWHM\_HA & (220-223) & I4 & $\rm{km}$ $\rm{s^{-1}}$  & FWHM of \ha\\
40 & FWHM\_HA\_UPP & (225-228) & I4 & $\rm{km}$ $\rm{s^{-1}}$  & Upper uncertainty of FWHM of \ha\\
41 & FWHM\_HA\_LOW & (230-233) & I4 & $\rm{km}$ $\rm{s^{-1}}$  & Lower uncertainty of FWHM of \ha\\
42 & EW\_HA & (235-237) & I3 & \AA & Rest-frame EW of \ha\\
43 & EW\_HA\_UPP & (239-240) & I2 & \AA & Upper uncertainty of EW of \ha\\
44 & EW\_HA\_LOW & (242-243) & I2 & \AA & Lower uncertainty of EW of \ha\\
45 & AS\_HA & (245-253) & E9.2 & \nodata & Asymmetry of the double Gaussian fit profile of \ha\\
46 & KURT\_HA & (255-258) & F4.2 & \nodata & Kurtosis of the double Gaussian fit profile of \ha\\
47 & FWHM\_FE II & (260-264) & F5.0 & $\rm{km}$ $\rm{s^{-1}}$ & FWHM of the kernel Gaussian used to broaden the \fetwo\ template\\
48 & EW\_FE II & (266-273) & E8.2 & \AA & Rest-frame EW of optical band \fetwo\ as defined by \cite{1992ApJS...80..109B} \\
49 & EW\_FE II\_UPP & (275-282) & E8.2 & \AA & Upper uncertainty of EW of \fetwo\\
50 & EW\_FE II\_LOW & (284-291) & E8.2 & \AA & Lower uncertainty of EW of \fetwo\\
51 & LOGF$\lambda$5100 & (293-298) & F6.2 & $\rm{erg}\ \rm{s^{-1}} \rm{cm^{-2}}\rm{\AA^{-1}}$ & Flux density at rest-frame 5100 \AA\\
52 & LOGL5100 & (300-304) & F5.2 & $\rm{erg}$ $\rm{s^{-1}}$ $\rm \AA^{-1}$ & Monochromatic luminosity at rest-frame 5100 \AA\ based on $z_{\rm sys}$ \\
\enddata
\tablenotetext{a}{The emission-line peak based on the peak-fit value.}
\tablecomments{Data formatting used for the catalog. Asymmetry is defined here as the skewness of the Gaussian fits, i.e., a measure of the asymmetry of the distribution about its mean, $s = E(x - \mu)^3 / \sigma^3$, where $\mu$ ~is the mean of $x$, $\sigma$ ~is the standard deviation of $x$, and $E(t)$ is the expectation value. Kurtosis is the quantification of the "tails" of the Gaussian fits defined as $k = E(x - \mu)^4 / \sigma^4$. All of the GNIRS spectra and their best-fit models are available electronically at\dataset[https://datalab.noirlab.edu/gnirs\_dqs.php]{https://datalab.noirlab.edu/gnirs_dqs.php}.}
\end{deluxetable*}

\begin{deluxetable*}{llllll}
\tablecolumns{6}
\tabletypesize{\scriptsize}
\tablewidth{0pc}
\tablecaption{Column Headings for Supplemental Emission-Line Measurements \label{tab:supp}}
\tablehead
{
\colhead{Column} &
\colhead{Name} &
\colhead{Bytes} &
\colhead{Format} &
\colhead{Units} &
\colhead{Description}\\
\colhead{(1)} &
\colhead{(2)} &
\colhead{(3)} &
\colhead{(4)} &
\colhead{(5)} &
\colhead{(6)}
}
\startdata
1 & OBJ & (1-24) & A24 & \nodata & SDSS object designation\\
2 & LC\_HD & (26-30) & I5 & \AA & \hd\ observed-frame wavelength\tablenotemark{a} \\
3 & LC\_HD\_UPP & (32-33) & I2 & \AA & Upper uncertainty for the line peak of \hd\\
4 & LC\_HD\_LOW & (35-36) & I2 & \AA & Lower uncertainty for the line peak of \hd\\
5 & FWHM\_HD & (38-41) & I4 & $\rm{km}$ $\rm{s^{-1}}$  & FWHM of \hd\\
6 & FWHM\_HD\_UPP & (43-45) & I3 & $\rm{km}$ $\rm{s^{-1}}$  & Upper uncertainty of FWHM of \hd\\
7 & FWHM\_HD\_LOW & (47-49) & I3 & $\rm{km}$ $\rm{s^{-1}}$  & Lower uncertainty of FWHM of \hd\\
8 & EW\_HD & (51-52) & I2 & \AA & Rest-frame EW of \hd\\
9 & EW\_HD\_UPP & (54-55) & I2 & \AA & Upper uncertainty of EW of \hd\\
10 & EW\_HD\_LOW & (57-58) & I2 & \AA & Lower uncertainty of EW of \hd\\
11 & AS\_HD & (60-68) & E9.2 & \nodata & Asymmetry of the double Gaussian fit profile of \hd\\
12 & KURT\_HD & (70-73) & F4.2 & \nodata & Kurtosis of the double Gaussian fit profile of \hd\\
13 & LC\_HG & (75-79) & I5 & \AA & \hg\ observed-frame wavelength\tablenotemark{a} \\
14 & LC\_HG\_UPP & (81-82) & I2 & \AA & Upper uncertainty for the line peak of \hg\\
15 & LC\_HG\_LOW & (84-85) & I2 & \AA & Lower uncertainty for the line peak of \hg\\
16 & FWHM\_HG & (87-90) & I4 & $\rm{km}$ $\rm{s^{-1}}$  & FWHM of \hg\\
17 & FWHM\_HG\_UPP & (92-95) & I4 & $\rm{km}$ $\rm{s^{-1}}$  & Upper uncertainty of FWHM of \hg\\
18 & FWHM\_HG\_LOW & (97-100) & I4 & $\rm{km}$ $\rm{s^{-1}}$  & Lower uncertainty of FWHM of \hg\\
19 & EW\_HG & (102-103) & I2 & \AA & Rest-frame EW of \hg\\
20 & EW\_HG\_UPP & (105-106) & I2 & \AA & Upper uncertainty of EW of \hg\\
21 & EW\_HG\_LOW & (108-109) & I2 & \AA & Lower uncertainty of EW of \hg\\
22 & AS\_HG & (111-119) & E9.2 & \nodata & Asymmetry of the double Gaussian fit profile of \hg\\
23 & KURT\_HG & (121-124) & F4.2 & \nodata & Kurtosis of the double Gaussian fit profile of \hg\\
24 & LC\_O II\tablenotemark{b} & (126-130) & I5 & \AA & \otwo\ observed-frame wavelength\tablenotemark{a} \\
25 & LC\_O II\_UPP & (132-133) & I2 & \AA & Upper uncertainty for the line peak of \otwo\\
26 & LC\_O II\_LOW & (135-136) & I2 & \AA & Lower uncertainty for the line peak of \otwo\\
27 & FWHM\_O II & (138-141) & I4 & $\rm{km}$ $\rm{s^{-1}}$  & FWHM of \otwo\\
28 & FWHM\_O II\_UPP & (143-147) & I5 & $\rm{km}$ $\rm{s^{-1}}$  & Upper uncertainty of FWHM of \otwo\\
29 & FWHM\_O II\_LOW & (149-152) & I4 & $\rm{km}$ $\rm{s^{-1}}$  & Lower uncertainty of FWHM of \otwo\\
30 & EW\_O II & (154-155) & I2 & \AA & Rest-frame EW of \otwo\\
31 & EW\_O II\_UPP & (157-158) & I2 & \AA & Upper uncertainty of EW of \otwo\\
32 & EW\_O II\_LOW & (160-161) & I2 & \AA & Lower uncertainty of EW of \otwo\\
33 & AS\_O II & (163-171) & E9.2 & \nodata & Asymmetry of the double Gaussian fit profile of \otwo\\
34 & KURT\_O II & (173-176) & F4.2 & \nodata & Kurtosis of the double Gaussian fit profile of \otwo\\
35 & LC\_NE III\tablenotemark{c} & (178-182) & I5 & \AA & \nethree\ observed-frame wavelength\tablenotemark{a} \\
36 & LC\_NE III\_UPP & (184-185) & I2 & \AA & Upper uncertainty for the line peak of \nethree\\
37 & LC\_NE III\_LOW & (187-188) & I2 & \AA & Lower uncertainty for the line peak of \nethree\\
38 & FWHM\_NE III & (190-193) & I4 & $\rm{km}$ $\rm{s^{-1}}$  & FWHM of \nethree\\
39 & FWHM\_NE III\_UPP & (195-198) & I4 & $\rm{km}$ $\rm{s^{-1}}$  & Upper uncertainty of FWHM of \nethree\\
40 & FWHM\_NE III\_LOW & (200-203) & I4 & $\rm{km}$ $\rm{s^{-1}}$  & Lower uncertainty of FWHM of \nethree\\
41 & EW\_NE III & (205-206) & I2 & \AA & Rest-frame EW of \nethree\\
42 & EW\_NE III\_UPP & (208-209) & I2 & \AA & Upper uncertainty of EW of \nethree\\
43 & EW\_NE III\_LOW & (211-212) & I2 & \AA & Lower uncertainty of EW of \nethree\\
44 & AS\_NE III & (214-222) & E9.2 & \nodata & Asymmetry of the double Gaussian fit profile of \nethree\\
45 & KURT\_NE III & (224-227) & F4.2 & \nodata & Kurtosis of the double Gaussian fit profile of \nethree\\
\enddata
\tablenotetext{a}{The emission-line peak based on the peak-fit value.}
\tablenotetext{b}{\otwo ~$\lambda$ 3727}
\tablenotetext{c}{\nethree ~$\lambda$ 3870}
\tablecomments{Data formatting used for the supplemental measurements in the supplemental features catalog.}
\end{deluxetable*}

\begin{deluxetable*}{llllll}
\tablecolumns{6}
\tabletypesize{\scriptsize}
\tablewidth{0pc}
\tablecaption{Column Headings for Gaussian Parameters of Emission-Line Profiles \label{tab:gauss}}
\tablehead
{
\colhead{Column} &
\colhead{Name} &
\colhead{Bytes} &
\colhead{Format} &
\colhead{Units} &
\colhead{Description}\\
\colhead{(1)} &
\colhead{(2)} &
\colhead{(3)} &
\colhead{(4)} &
\colhead{(5)} &
\colhead{(6)}
}
\startdata
1 & OBJ & (1-24) & A24 & \nodata & SDSS object designation\\
2 & MG II\_LAM\_PEAK\_NARROW & (26-29) & I4 & \AA & Narrow \mgtwo\  peak\tablenotemark{a} \\
3 & MG II\_STD\_NARROW & (31-32) & I2 & \AA & Narrow \mgtwo\  width\\
4 & MG II\_F\_LAM\_NARROW & (34-37) & I4 & $\rm{erg}\ \rm{s^{-1}} \rm{cm^{-2}}\rm{\AA^{-1}}$ & Narrow \mgtwo\  normalization\\
5 & MG II\_LAM\_PEAK\_BROAD & (39-42) & I4 & \AA & Broad \mgtwo\  peak\tablenotemark{a} \\
6 & MG II\_STD\_BROAD & (44-47) & I4 & \AA & Broad \mgtwo\  width\\
7 & MG II\_F\_LAM\_BROAD & (49-52) & I4 & $\rm{erg}\ \rm{s^{-1}} \rm{cm^{-2}}\rm{\AA^{-1}}$ & Broad \mgtwo\  normalization\\
8 & O II\_LAM\_PEAK\_NARROW & (54-57) & I4 & \AA & Narrow \otwo\  peak\tablenotemark{a} \\
9 & O II\_STD\_NARROW & (59-60) & I2 & \AA & Narrow \otwo\  width\\
10 & O II\_F\_LAM\_NARROW & (62-65) & I4 & $\rm{erg}\ \rm{s^{-1}} \rm{cm^{-2}}\rm{\AA^{-1}}$ & Narrow \otwo\  normalization\\
11 & O II\_LAM\_PEAK\_BROAD & (67-70) & I4 & \AA & Broad \otwo\  peak\tablenotemark{a} \\
12 & O II\_STD\_BROAD & (72-75) & I4 & \AA & Broad \otwo\  width\\
13 & O II\_F\_LAM\_BROAD & (77-78) & I2 & $\rm{erg}\ \rm{s^{-1}} \rm{cm^{-2}}\rm{\AA^{-1}}$ & Broad \otwo\  normalization\\
14 & NE III\_LAM\_PEAK\_NARROW & (80-83) & I4 & \AA & Narrow \nethree\  peak\tablenotemark{a} \\
15 & NE III\_STD\_NARROW & (85-86) & I2 & \AA & Narrow \nethree\  width\\
16 & NE III\_F\_LAM\_NARROW & (88-89) & I2 & $\rm{erg}\ \rm{s^{-1}} \rm{cm^{-2}}\rm{\AA^{-1}}$ & Narrow \nethree\  normalization\\
17 & NE III\_LAM\_PEAK\_BROAD & (91-94) & I4 & \AA & Broad \nethree\  peak\tablenotemark{a} \\
18 & NE III\_STD\_BROAD & (96-99) & I4 & \AA & Broad \nethree\  width\\
19 & NE III\_F\_LAM\_BROAD & (101-102) & I2 & $\rm{erg}\ \rm{s^{-1}} \rm{cm^{-2}}\rm{\AA^{-1}}$ & Broad \nethree\  normalization\\
20 & HD\_LAM\_PEAK\_NARROW & (104-107) & I4 & \AA & Narrow \hd\  peak\tablenotemark{a} \\
21 & HD\_STD\_NARROW & (109-110) & I2 & \AA & Narrow \hd\  width\\
22 & HD\_F\_LAM\_NARROW & (112-113) & I2 & $\rm{erg}\ \rm{s^{-1}} \rm{cm^{-2}}\rm{\AA^{-1}}$ & Narrow \hd\  normalization\\
23 & HD\_LAM\_PEAK\_BROAD & (115-118) & I4 & \AA & Broad \hd\  peak\tablenotemark{a} \\
24 & HD\_STD\_BROAD & (120-123) & I4 & \AA & Broad \hd\  width\\
25 & HD\_F\_LAM\_BROAD & (125-127) & I3 & $\rm{erg}\ \rm{s^{-1}} \rm{cm^{-2}}\rm{\AA^{-1}}$ & Broad \hd\  normalization\\
26 & HG\_LAM\_PEAK\_NARROW & (129-132) & I4 & \AA & Narrow \hg\  peak\tablenotemark{a} \\
27 & HG\_STD\_NARROW & (134-135) & I2 & \AA & Narrow \hg\  width\\
28 & HG\_F\_LAM\_NARROW & (137-139) & I3 & $\rm{erg}\ \rm{s^{-1}} \rm{cm^{-2}}\rm{\AA^{-1}}$ & Narrow \hg\  normalization\\
29 & HG\_LAM\_PEAK\_BROAD & (141-144) & I4 & \AA & Broad \hg\  peak\tablenotemark{a} \\
30 & HG\_STD\_BROAD & (146-149) & I4 & \AA & Broad \hg\  width\\
31 & HG\_F\_LAM\_BROAD & (151-153) & I3 & $\rm{erg}\ \rm{s^{-1}} \rm{cm^{-2}}\rm{\AA^{-1}}$ & Broad \hg\  normalization\\
32 & HB\_LAM\_PEAK\_NARROW & (155-158) & I4 & \AA & Narrow \hb\  peak\tablenotemark{a} \\
33 & HB\_STD\_NARROW & (160-162) & I3 & \AA & Narrow \hb\  width\\
34 & HB\_F\_LAM\_NARROW & (164-166) & I3 & $\rm{erg}\ \rm{s^{-1}} \rm{cm^{-2}}\rm{\AA^{-1}}$ & Narrow \hb\  normalization\\
35 & HB\_LAM\_PEAK\_BROAD & (168-171) & I4 & \AA & Narrow \hb\  peak\tablenotemark{a} \\
36 & HB\_STD\_BROAD & (173-175) & I3 & \AA & Broad \hb\  width\\
37 & HB\_F\_LAM\_BROAD & (177-179) & I3 & $\rm{erg}\ \rm{s^{-1}} \rm{cm^{-2}}\rm{\AA^{-1}}$ & Broad \hb\  normalization\\
38 & O III\_1\_LAM\_PEAK\_NARROW & (181-184) & I4 & \AA & Narrow \othree\ 4959\AA\  peak\tablenotemark{a} \\
39 & O III\_1\_STD\_NARROW & (186-187) & I2 & \AA & Narrow \othree\ 4959\AA\  width\\
40 & O III\_1\_F\_LAM\_NARROW & (189-191) & I3 & $\rm{erg}\ \rm{s^{-1}} \rm{cm^{-2}}\rm{\AA^{-1}}$ & Narrow \othree\ 4959\AA\  normalization\\
41 & O III\_1\_LAM\_PEAK\_BROAD & (193-196) & I4 & \AA & Broad \othree\ 4959\AA\  peak\tablenotemark{a} \\
42 & O III\_1\_STD\_BROAD & (198-200) & I3 & \AA & Broad \othree\ 4959\AA\  width\\
43 & O III\_1\_F\_LAM\_BROAD & (202-204) & I3 & $\rm{erg}\ \rm{s^{-1}} \rm{cm^{-2}}\rm{\AA^{-1}}$ & Broad \othree\ 4959\AA\  normalization\\
44 & O III\_2\_LAM\_PEAK\_NARROW & (206-209) & I4 & \AA & Narrow \othree\ 5007\AA\  peak\tablenotemark{a} \\
45 & O III\_2\_STD\_NARROW & (211-212) & I2 & \AA & Narrow \othree\ 5007\AA\  width\\
46 & O III\_2\_F\_LAM\_NARROW & (214-216) & I3 & $\rm{erg}\ \rm{s^{-1}} \rm{cm^{-2}}\rm{\AA^{-1}}$ & Narrow \othree\ 5007\AA\  normalization\\
47 & O III\_2\_LAM\_PEAK\_BROAD & (218-221) & I4 & \AA & Broad \othree\ 5007\AA\  peak\tablenotemark{a} \\
48 & O III\_2\_STD\_BROAD & (223-225) & I3 & \AA & Broad \othree\ 5007\AA\  width\\
49 & O III\_2\_F\_LAM\_BROAD & (227-229) & I3 & $\rm{erg}\ \rm{s^{-1}} \rm{cm^{-2}}\rm{\AA^{-1}}$ & Broad \othree\ 5007\AA\  normalization\\
50 & HA\_LAM\_PEAK\_NARROW & (231-234) & I4 & \AA & Narrow \ha\  peak\tablenotemark{a} \\
51 & HA\_STD\_NARROW & (236-238) & I3 & \AA & Narrow \ha\  width\\
52 & HA\_F\_LAM\_NARROW & (240-243) & I4 & $\rm{erg}\ \rm{s^{-1}} \rm{cm^{-2}}\rm{\AA^{-1}}$ & Narrow \ha\  normalization\\
53 & HA\_LAM\_PEAK\_BROAD & (245-248) & I4 & \AA & Broad \ha\  peak\tablenotemark{a} \\
54 & HA\_STD\_BROAD & (250-252) & I3 & \AA & Broad \ha\  width\\
55 & HA\_F\_LAM\_BROAD & (254-256) & I3 & $\rm{erg}\ \rm{s^{-1}} \rm{cm^{-2}}\rm{\AA^{-1}}$ & Broad \ha\  normalization\\
\enddata
\tablenotetext{a}{The Gaussian profile peak based on the peak-fit value.}
\tablecomments{Independent Gaussian feature fit parameters for each emission line that was fit with both a narrow and broad Gaussian profile.}
\end{deluxetable*}

\subsection{\cfour\ Emission-Line Measurements}

M17 and D20 found that the accuracy and precision of a source's UV-based redshift can be significantly improved when regressed against the FWHM and EW of its \cfour\ line as well as the UV continuum luminosity at a rest-frame wavelength of 1350\AA\ ($L_{\rm 1350}$).\footnote{Objects with redshifts $z < 1.65$ had $L_{\rm 1350}$ values extrapolated from $L_{\rm 3000}$ assuming a continuum power-law of the form $f_\nu$ $\propto$ $\nu^{0.5}$ \citep[e.g.,][]{2001AJ....122..549V}.} The \cfour\ emission line has been measured in the SDSS spectrum of each GNIRS-DQS source using the same fitting approach outlined in D20, which closely follows the methods utilized in both M21 and this work; the \cfour\ emission-line properties of all the GNIRS-DQS sources appear in \citeauthor{dix23}.

\section{Correcting UV-Based Redshifts} \label{sec:data}

Our aim is to derive corrections that, on average, shift the velocity offsets of the UV-based redshifts as close as possible to a velocity offset of zero $\rm{km~s^{-1}}$ from $z_{\rm sys}$ based on the \othree\ $\lambda5007$ line. We derive this correction by applying a regression analysis to a calibration sample of 121 sources from GNIRS-DQS as described below.

The sample used for this analysis is a subset of the augmented GNIRS-DQS sample described in Section~\ref{sec:sample}. Starting with the 260 GNIRS-DQS sources with useful NIR spectra, we include only the 220 objects with \othree\ rest-frame EW measurements greater than \hbox{1 \AA} that can provide the most accurate values of $z_{\rm sys}$ (see Figure~\ref{fig:o3ew}); i.e., 40 sources whose $z_{\rm sys}$ values were based on either \mgtwo ~or \hb\ were removed. We then remove 52 broad absorption line (BAL) quasars, identified as such based on criteria outlined in their respective source catalogs using ``balnicity" index and visual inspection \citep[see,][]{2009ApJ...692..758G,2011ApJS..194...45S,2017A&A...597A..79P,2018A&A...613A..51P,2020ApJS..250....8L}, as the BAL troughs often degrade measurements of the EW and FWHM for \cfour, which are of primary importance for our regression analysis \citep[e.g.,][]{1991ApJ...382..416B,2009ApJ...692..758G}. We also remove 17 radio-loud (RL) quasars (having $R > 100$; see Figure~\ref{fig:RL}), one of which, SDSS J114705.24+083900.6, is also classified as a BAL quasar, due to potential continuum boosting, which may affect both EW (\cfour) and $L_{1350}$ measurements \citep[e.g.,][]{2011ApJ...726...20M}.

Two additional sources, SDSS J073132.18+461347.0, and SDSS J141617.38+264906.1, were excluded due to the inability to measure the \cfour\ line reliably (see \citeauthor{dix23}). Finally, 29 objects were removed from the sample due to additional visual inspection and a lack of a prominent \othree ~$\lambda$5007 emission line. The result of this selection process is a calibration sample of 121 sources, presented in Table~\ref{tab:zvs}, which is a representative sample of optically selected quasars (see Section~\ref{sec:sample}) used to derive prescriptions for correcting UV-based redshifts through linear regression analysis.

\begin{deluxetable*}{lccrcccr}
\tablecolumns{9}
\tabletypesize{\scriptsize}
\tablewidth{0pc}
\tablecaption{Redshifts and Velocity Offsets \label{tab:zvs}}
\tablehead
{
\colhead{} &
\colhead{} &
\colhead{} &
\colhead{$\Delta v_i$} &
\colhead{} &
\colhead{$\Delta v_i$} &
\colhead{} &
\colhead{$\Delta v_i$} \\
\colhead{Quasar} &
\colhead{$z_{\rm sys}$\tablenotemark{a}} &
\colhead{$z_{\rm C\,IV}$\tablenotemark{b} } &
\colhead{($\rm{km~s^{-1}}$)} &
\colhead{$z_{\rm HW10}$\tablenotemark{c} } &
\colhead{($\rm{km~s^{-1}}$)} &
\colhead{$z_{\rm Pipe}$\tablenotemark{d} } &
\colhead{($\rm{km~s^{-1}}$)}
}
\startdata
SDSS J001018.88+280932.5 &1.613 &1.611 &-230   & \nodata     &\nodata      & 1.612 & -110 \\
SDSS J001453.20+091217.6 &2.340 &2.326 &-1250 &2.344 &340 &2.308 &-2840   \\
SDSS J001914.46+155555.9 &2.267 &2.263 &-370   &2.276 &830 &2.271 &350      \\
SDSS J002634.46+274015.5 &2.247 &2.243 &-340   &2.247 &50 &2.267 &1850    \\
SDSS J003001.11$-$015743.5 &1.588 &1.579 &-1030 &1.590  &200  &1.582 & -710 \\
SDSS J003416.61+002241.1 &1.631 &1.626 &-560  & 1.630 & -50   &1.627 &-410    \\
SDSS J004710.48+163106.5 &2.192 &2.162 &-2770 &\nodata  & \nodata  &2.165 & -2490 \\
SDSS J004719.71+014813.9 &1.591 &1.588 &-340   &1.590 & -130 &1.590 &-130      \\
SDSS J005233.67+014040.8 &2.309 &2.295 &-1250 &2.305 &-370  &2.291 &-1620\\
SDSS J005307.71+191022.7 &1.598 &1.581 &-1940 &1.585 &-1460 &1.583 &-1680
\enddata
\tablenotetext{a}{Redshifts determined from the \othree ~$\lambda_{\rm peak}$ ~as described in M21.}
\tablenotetext{b}{Redshifts determined from the \cfour ~$\lambda_{\rm peak}$ values given in \citeauthor{dix23}.}
\tablenotetext{c}{Acquired from HW10 and/or from P. Hewett, priv. comm.}
\tablenotetext{d}{Acquired from Lyke et al. (\citeyear{2020ApJS..250....8L})}
\tablecomments{Complete table of 121 sources appears in the electronic version.}
\end{deluxetable*}

The redshift corrections are performed on redshifts obtained using three separate techniques: 1) measurements of the observed-frame wavelength of the peak of the \cfour ~emission line, 2) HW10 redshifts (P.~Hewett, priv. comm.), and 3) SDSS Pipeline redshifts \citep[][Table D1, column 29 ``Z\_PIPE"]{2012AJ....144..144B}. The HW10 redshifts are notable as they already have a primary redshift correction applied.

\begin{figure*}[]
\includegraphics[scale=0.58]{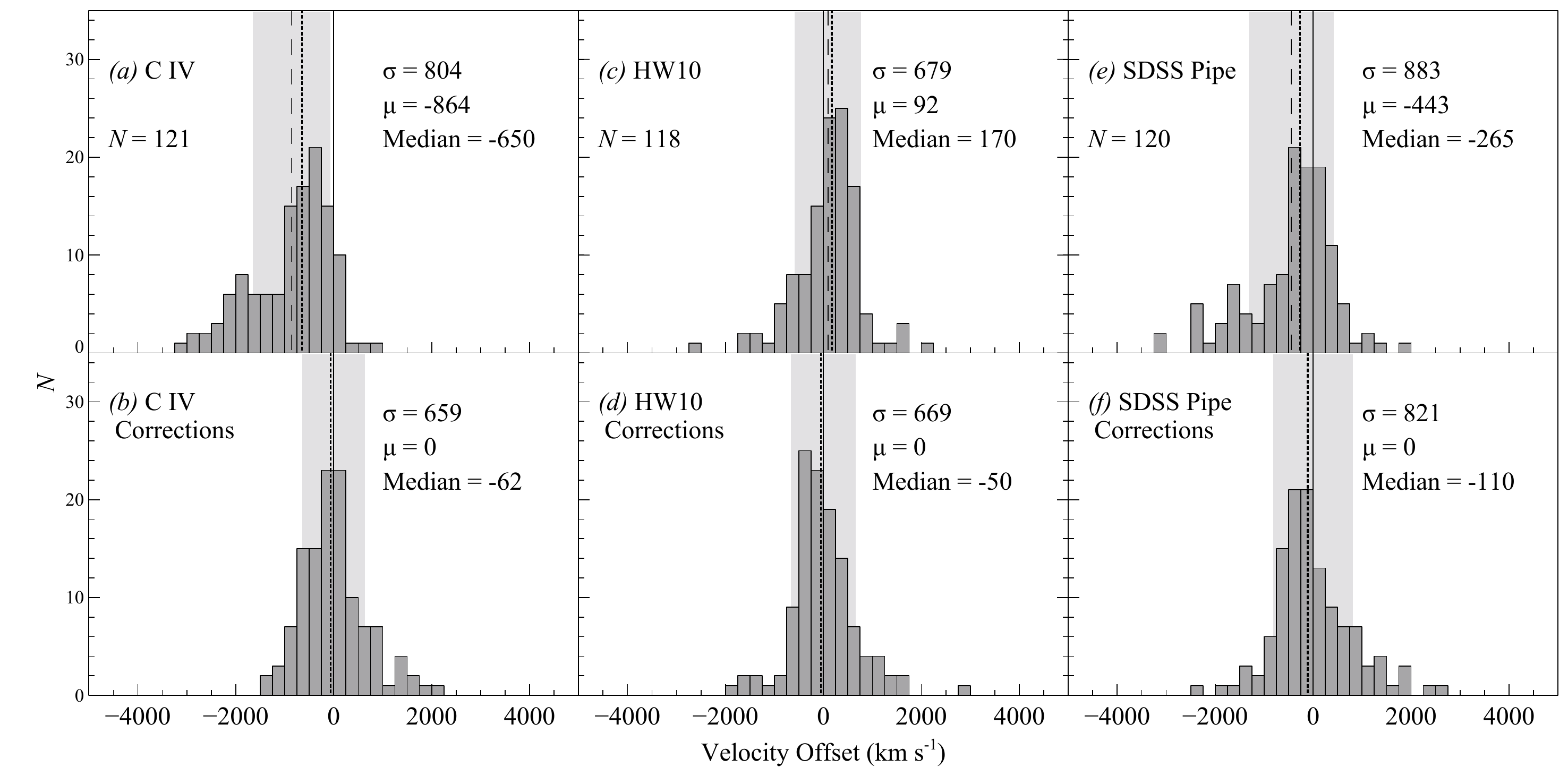}\caption{Velocity offsets relative to $z_{\rm{sys}}$ before (panels \textit{a}, \textit{c}, and \textit{e}) and after (panels \textit{b}, \textit{d}, and \textit{f}) the corrections using the linear regression coefficients given in Table~\ref{tab:coeff1}. The standard deviation (shaded region), mean (dashed line), median (dotted line), and zero velocity offset (solid line) are marked in each panel. SDSS J090247.57+304120.7 does not appear on the SDSS Pipe panels due to its erroneous redshift. The mean ($\mu$), median, and standard deviation ($\sigma$) given in units of$~\rm{km~s^{-1}}$, are noted in each panel. The number of sources used in each regression analysis ($N$) is given in the upper panels.}
\label{fig:154}
\end{figure*}

The principal metric under investigation in this work is the initial velocity offset ($\Delta v_i$) between each of the aforementioned three UV-based redshifts ($z_{\rm meas}$) and the $z_{\rm sys}$ value of a source determined from its \othree\ $\lambda5007$ emission line by measuring the line peak in each spectrum, which is presented in Table~\ref{tab:obj}. This offset is computed using the following equation (see D20):

\begin{equation} \label{eq:voff}
\Delta v_{i} = c \left( \frac{z_{\rm meas}-z_{\rm sys}} {1+z_{\rm sys}} \right).
\end{equation} \\
These initial velocity offset values are presented in Table~\ref{tab:zvs} and are shown in the top panels of Figure~\ref{fig:154}.

As shown in Table~\ref{tab:zvs}, there is one source, SDSS J090247.57+304120.7, where the SDSS Pipeline produces an erroneous redshift, resulting in an unrealistically high velocity offset of \hbox{$|\Delta \it{v_i}| > \rm{16000}~\rm{km~s^{-1}}$}, while the velocity offsets for this source from the \cfour\ and HW10 methods yield values that are only $-170$ and $+70~\rm{km~s^{-1}}$, respectively. As a result, this source is excluded from the SDSS Pipeline analysis, but is retained in the \cfour\ and HW10 analyses.

It is known that the \cfour\ velocity offsets are correlated with the \cfour\ FWHM \citep[e.g.,][]{2017MNRAS.465.2120C}. However, as was discovered by M17 and D20, additional corrections to the velocity offsets can be obtained by including two additional parameters: EW(\cfour) and $L_{1350}$. In this work, we confirm that all three parameters, FWHM(\cfour), EW(\cfour), and $L_{1350}$, are required for obtaining the best corrections for the velocity offsets in the following manner:

\begin{multline} \label{eq:vcorr}
\Delta v_{\rm corr} ~(\rm{km~s^{-1}}) = \alpha \rm{logFWHM_{C IV}} ~(\rm{km~s^{-1}})\\+\ \beta \rm{logEW_{C IV}} ~(\AA)\\+\ \gamma \rm{log\it{L}_{\rm 1350}}\ (10^{-17}\ erg\ s^{-1} \AA^{-1})
\end{multline} \\ where $\Delta v_{\rm corr}$ is the velocity offset we subtract from the initial velocity offset calculated using Equation~\ref{eq:voff}.

The final, post-correction velocity offset, $\Delta v_f = \Delta v_i - \Delta v_{\rm corr}$, is displayed in the bottom panels of Figure~\ref{fig:154}. Since the goal of Equation~\ref{eq:vcorr} is to eliminate the velocity offsets, then, by definition, the mean ($\mu$) of $\Delta v_i - \Delta v_{corr}$ is zero. This $\Delta v_{\rm corr}$ value is used to obtain a revised $z_{\rm{sys}}$ prediction by adjusting the initially measured redshift of a quasar. From Equation~\ref{eq:voff}, solving for $z_{\rm meas}$, and substituting $z_{\rm meas}$ = $z_{\rm sys}$ and $v_{\rm corr}$ = $v_{i}$, we get

\begin{equation} \label{eq:znew}
		z_{\rm rev} = z_{\rm meas} + \frac{\Delta v_{\rm corr} (1 + z_{\rm meas})}{c}
\end{equation} where $z_{\rm rev}$ is the revised, more accurate, and more precise redshift.

Starting with our 121-source calibration sample, we run linear regressions using the three parameters defined in Equation~\ref{eq:vcorr}. The results provide the $\Delta v_{\rm corr}$ values from Equation~\ref{eq:vcorr} that are subtracted from the initial velocity offsets of the sources (from Table~\ref{tab:zvs}).

Distributions of the $\Delta v_i$ and $\Delta v_f$ values are plotted in the top and bottom panels in Figure~\ref{fig:154}, respectively. We observe that the \cfour -based $\Delta v_i$ values are skewed toward negative values (blueshifts) with a mean velocity offset of $\mu = -864~\rm{km~s^{-1}}$, and a standard deviation of $\sigma = 804~\rm{km~s^{-1}}$. The SDSS Pipeline-based $\Delta v_i$ values have a considerably smaller negative initial velocity offset of $\mu = -443~\rm{km~s^{-1}}$, yet a larger standard deviation of $\sigma = 883~\rm{km~s^{-1}}$.  As expected, the HW10-based $\Delta v_i$ values show a mean initial velocity offset much closer to zero ($\mu = 92~\rm{km~s^{-1}}$), with the standard deviation being notably smaller than that of the \cfour -based $\Delta v_i$ values ($\sigma = 679~\rm{km~s^{-1}}$). Despite the improvements demonstrated by the HW10-based values, we are able to use our regression analysis to improve on UV-based redshift determinations further, as shown below.

As explained above, our redshift corrections yield mean $\Delta v_f$ values of zero $\rm{km~s^{-1}}$ using all three UV-based methods (see the bottom panels of Figure~\ref{fig:154}). The standard deviations ($\sigma$) of the $\Delta v_f$ values, on the other hand, are reduced by $\sim18\%$, $\sim2\%$, and $\sim7\%$ for the \cfour , HW10, and SDSS Pipeline methods, respectively, with respect to the measured $\Delta v_i$ values. The median velocity offsets are also reduced significantly for all three methods. The linear regression coefficients (Equation~\ref{eq:vcorr}) used to achieve these corrections are presented in Table~\ref{tab:coeff1}. The uncertainties on the coefficients, shown in Table~\ref{tab:coeff1}, from the regression defined in Equation~\ref{eq:vcorr} stem directly from the mean squared error from the linear fit. These uncertainties (translated to $\sim700~\rm{km~s^{-1}}$ in velocity space; see Figure~\ref{fig:154}) are considerably larger than the intrinsic uncertainties associated with the \othree\ line wavelength calibration (on the order of $\sim10~\rm{km~s^{-1}}$; see M21) or the deviation of the \othree\ line from the systemic redshift ($-48~\rm{km~s^{-1}}$ systematic offset, and $56~\rm{km~s^{-1}}$ uncertainty; \citeauthor{2016ApJ...831....7S} \citeyear{2016ApJ...831....7S}); therefore we adopt the regression uncertainties as the most conservative error estimates. Table~\ref{tab:coeff1} also gives the \textit{t}-Value \citep[e.g.,][]{sheskin07} for confidence statistics in determining the importance of each parameter (see also D20), where \textit{t}-Values of $|t|\gtrsim2$ denote a strong correlation, with decreasing confidence as $t \rightarrow 0$.

\begin{deluxetable*}{cccrrr}
\tablecaption{Linear Regression Coefficients \label{tab:coeff1}}
\tablewidth{0pt}
\tablehead{
\colhead{UV-Based} & \colhead{Sample} & \colhead{Regression} & \colhead{Value} & \colhead{Error} & \colhead{$t$-Value} \\
\colhead{Redshift Method} & \colhead{Size} & \colhead{Coefficients} & \colhead{} & \colhead{} & \colhead{}
}
\decimalcolnumbers
\startdata
& & $\alpha$ & -2589 & 428 & -6.05 \\
\cfour & 121 & $\beta$ & 1156  & 292 & 3.95\\
& & $\gamma$ & 148 & 36 & 4.15 \\
\hline
& & $\alpha$ & -587 & 389& -1.51 \\
HW10   & 118 & $\beta$ & 283 & 261 & 1.09\\
& & $\gamma$ & 39  & 32 & 1.21 \\
\hline
& & $\alpha$ & 9371 &8557 & 1.10 \\
SDSS Pipe & 120 & $\beta$ & -3694 & 5851 & -0.63\\
& & $\gamma$ & -655  &715 &-0.92 \\
\enddata
\end{deluxetable*}

\begin{figure*}[]
\includegraphics[scale=0.68]{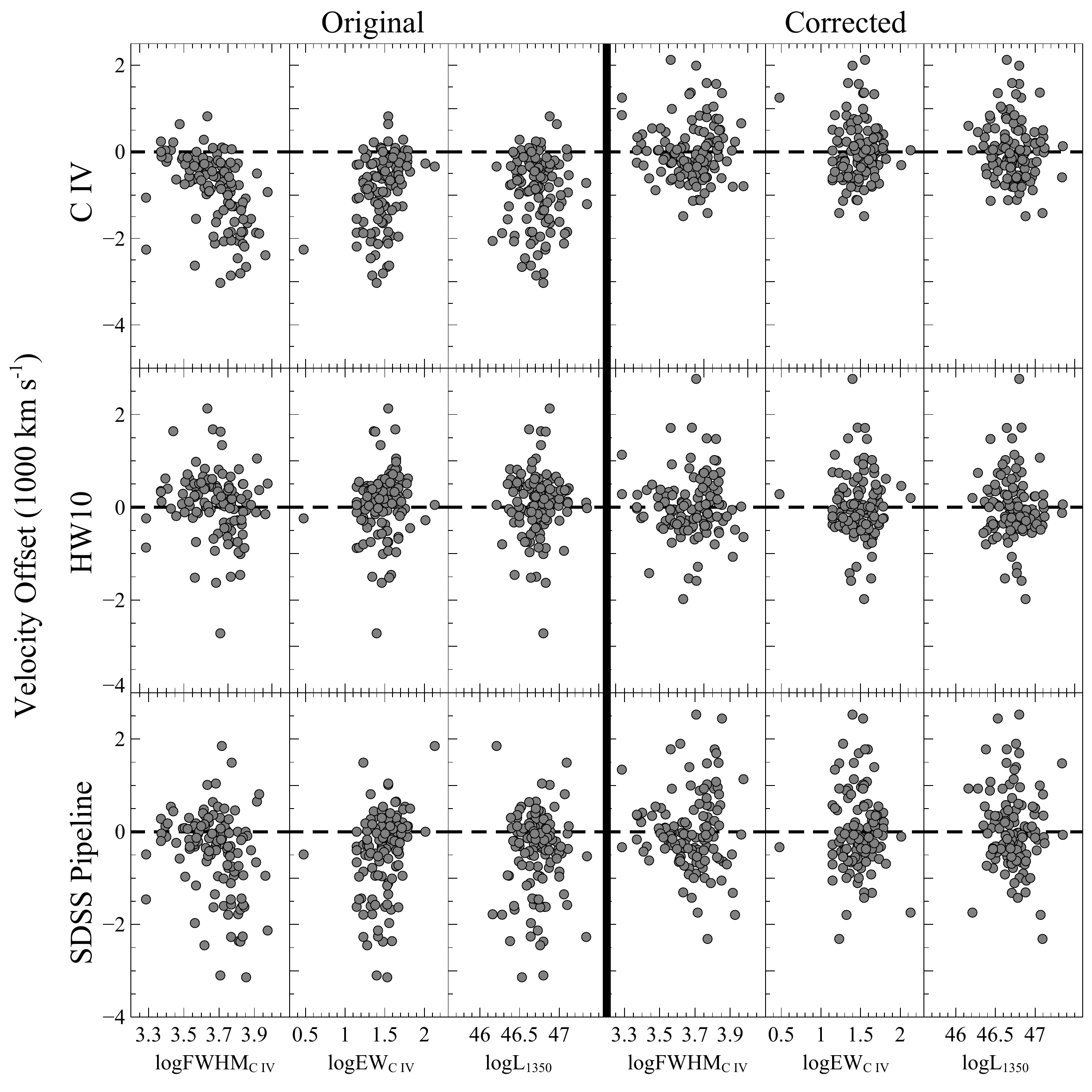}\caption{Residual velocity offsets with respect to $z_{\rm sys}$ before (three leftmost columns), and after (three rightmost columns), corrections are applied (see Equation~\ref{eq:vcorr}) against our regression parameters. SDSS J090247.57+304120.7 does not appear in the SDSS Pipeline panels due to its erroneously reported redshift, as discussed in Section~\ref{sec:data}.}
\label{fig:res}
\end{figure*}

Residuals of the 121 source sample both before and after our corrections are applied are presented in Figure~\ref{fig:res}. The residual distributions show the substantial reduction in the velocity offsets before and after each correction. The corrected velocity offsets for both the \cfour\ and HW10-based methods are closer to zero than the corrected velocity offsets for the SDSS Pipeline method.

The relatively small improvement in the mean velocity offset ($\mu$), and the standard deviation ($\sigma$), achieved for the HW10 method is likely related to the fact that our analysis constitutes a second-order correction to the one already employed by HW10; this result reinforces the general reliability of that method. The \cfour -based redshifts, while based solely on a \textit{single emission line}, provide a slightly smaller standard deviation than the HW10-based method when corrected using our regression analysis (see Figure~\ref{fig:154}). Finally, the SDSS Pipeline-based redshifts provide the least reliable results; in particular, we find that the SDSS Pipeline fails to produce a meaningful redshift for one out of 121 sources in our calibration sample (SDSS J090247.57+304120.7). Furthermore, DR16 redshifts \citep[][Table D1, column 10 ``$\rm{Z}\_\rm{QN}$"]{2020ApJS..250....8L} available for 101 sources from our sample provide significantly larger standard deviations than the SDSS Pipeline values both before and after the correction.

\begin{figure}[]
\includegraphics[scale=0.5]{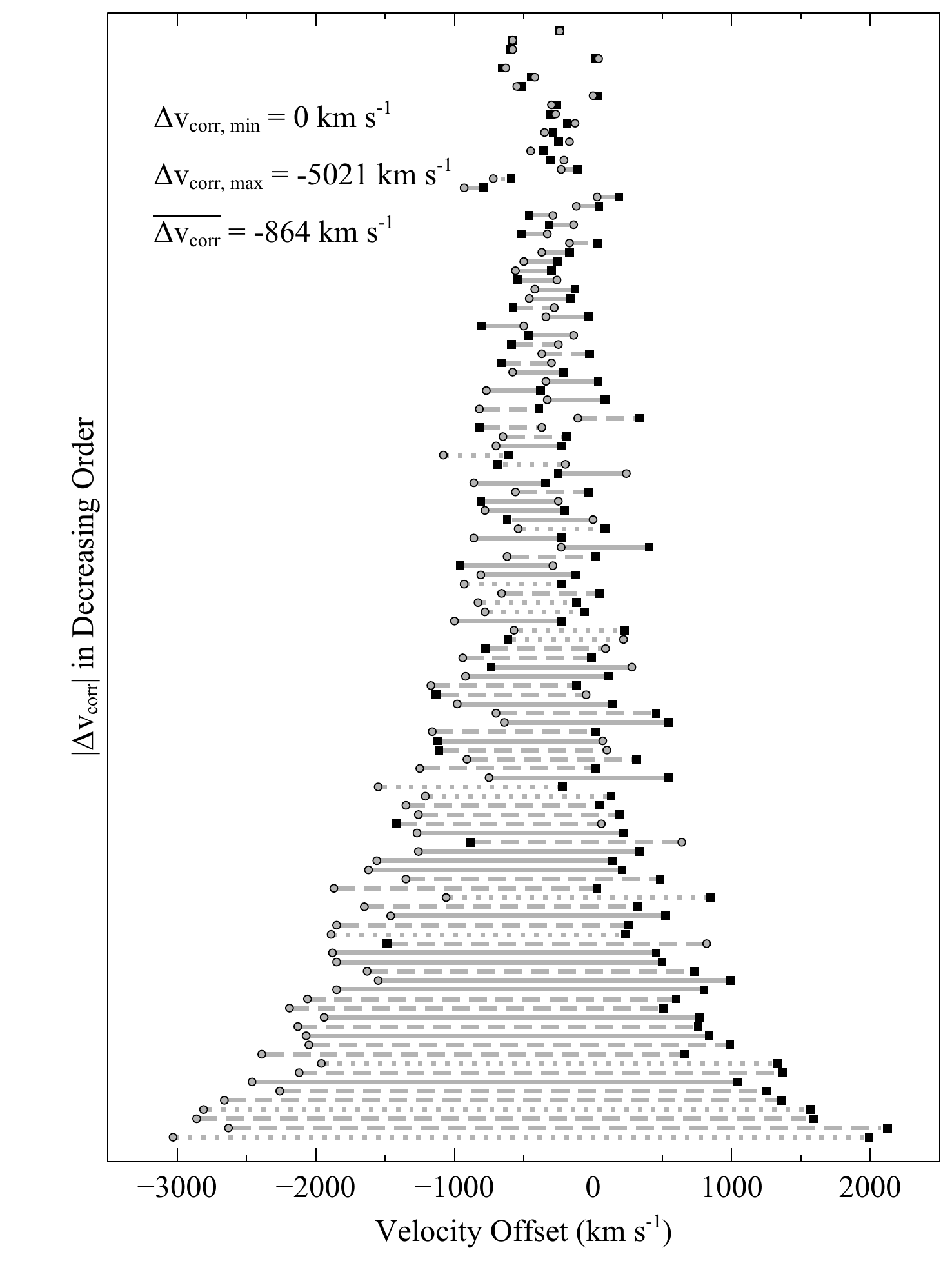}\caption{Initial velocity offsets ($\Delta v_i$; circles) compared to final velocity offsets ($\Delta v_f$; squares) for \cfour -based redshifts of the calibration sample of 121 sources. The lines connecting the initial and final velocity offsets are sorted from top to bottom by the absolute value of the velocity offset correction ($|\Delta v_{\rm corr}|$). Solid lines, dashed lines, and dotted lines refer to the following luminosity ranges: \hbox{$46.08 < $~log($L_{5100}$)~$ < 46.41$}, \hbox{$46.42 < $~log($L_{5100}$)~$ < 46.74$}, and \hbox{$46.75 < ~$log($L_{5100}$)~$ < 47.09$}, respectively. Fifteen sources have $|\Delta v_{\rm corr}| < 100~\rm{km~s^{-1}}$, 11, three, and one of which belong to the lowest, medium, and highest luminosity ranges, respectively. While the majority of the $\Delta v_i$ values, which are blueshifts, produce $\Delta v_f$ values with the opposite sign, we also see $\Delta v_i$ values which are redshifts that end up as blueshifts; however the overall effect of our regression analysis brings $\Delta v_f$ values closer to zero. We find no trend between $|\Delta v_{\rm corr}|$ and the monochromatic luminosity at rest-frame 5100 \AA. }
\label{fig:swing}
\end{figure}

For our calibration sample, the largest $\Delta v_i$ value is on the order of $\sim-3000~\rm{km~s^{-1}}$, with a few sources having velocity offsets within the \hbox{$-3000 ~\rm{km~s^{-1}}$ $\lesssim \Delta v_i \lesssim$ $-2000 ~\rm{km~s^{-1}}$} range. Values of this magnitude, while high, are not unexpected due to the kinematics associated with luminous, rapidly accreting quasars that can directly affect the \cfour\ emission line and cause large blueshifts \citep[e.g.,][]{1995ApJ...451..498M,2007ApJ...666..757S,2008ApJ...680..169S,2020ApJ...893...14D}. Panel $(a)$ of Figure~\ref{fig:154}, and to a lesser extent panel $(e)$, demonstrate the impact of this effect via the asymmetric distribution caused by this large range of velocity offsets.  Nevertheless, our method tends to correct even these large velocity offsets to more reasonable values as shown in Figures~\ref{fig:154} and~\ref{fig:swing}.

Our velocity offset distributions may appear more asymmetric with respect to previous studies of this kind. For example, Shen et al. (\citeyear{2011ApJS..194...45S}), who derived \cfour\ velocity offsets based on \mgtwo\ (from the same SDSS spectrum), provide more symmetric distributions than those presented in Figure~\ref{fig:154}. This may be a result of the much larger uncertainties in the determination of $z_{\rm sys}$ from \mgtwo\ that was used in that study, compared with the uncertainties associated with the \othree\ lines, which are up to an order of magnitude smaller (e.g., M21).

The results of our regression analysis, presented in Table~\ref{tab:coeff1}, provide considerably improved redshifts over the regression coefficients used by D20. When we employ the D20 regression coefficients on our calibration sample of 121 sources, we obtain standard deviations for the distributions of $\Delta v_f$ which are $\sim18\%$ larger for the HW10 method, $\sim32\%$ larger for the SDSS Pipeline method, and $\sim4\%$ larger for the \cfour -based redshifts compared to using the coefficients from Table~\ref{tab:coeff1}. In summary, considering the four basic observables associated with the \cfour\ emission line, one can derive the most accurate and precise prediction of the systemic redshift of a quasar using that sole emission line.

\subsection{Redshift and Luminosity Dependence \label{sec:regimes}}

Typically, redshifts are determined either spectroscopically or photometrically from multiple features (i.e., HW10 and the SDSS Pipeline). When some of these features are no longer available in a spectrum, our ability to determine the redshift is affected, and it is plausible that the initial velocity offsets depend also on source redshift. We search for such a dependence in our data by splitting our calibration sample into three redshift bins: $1.55 \lesssim z \lesssim 1.65$ (Bin 1), $2.10 \lesssim z \lesssim 2.40$ (Bin 2), and $3.20 \lesssim z \lesssim 3.50$ (Bin 3), which contain 37, 71, and 13 sources, respectively. These intervals ensure coverage of the \othree~$\lambda5007$ emission line in the $J$, $H$, or $K$ bands, respectively (see Section~\ref{sec:sample}).

\begin{figure*}[]
\centering
\includegraphics[scale=0.75]{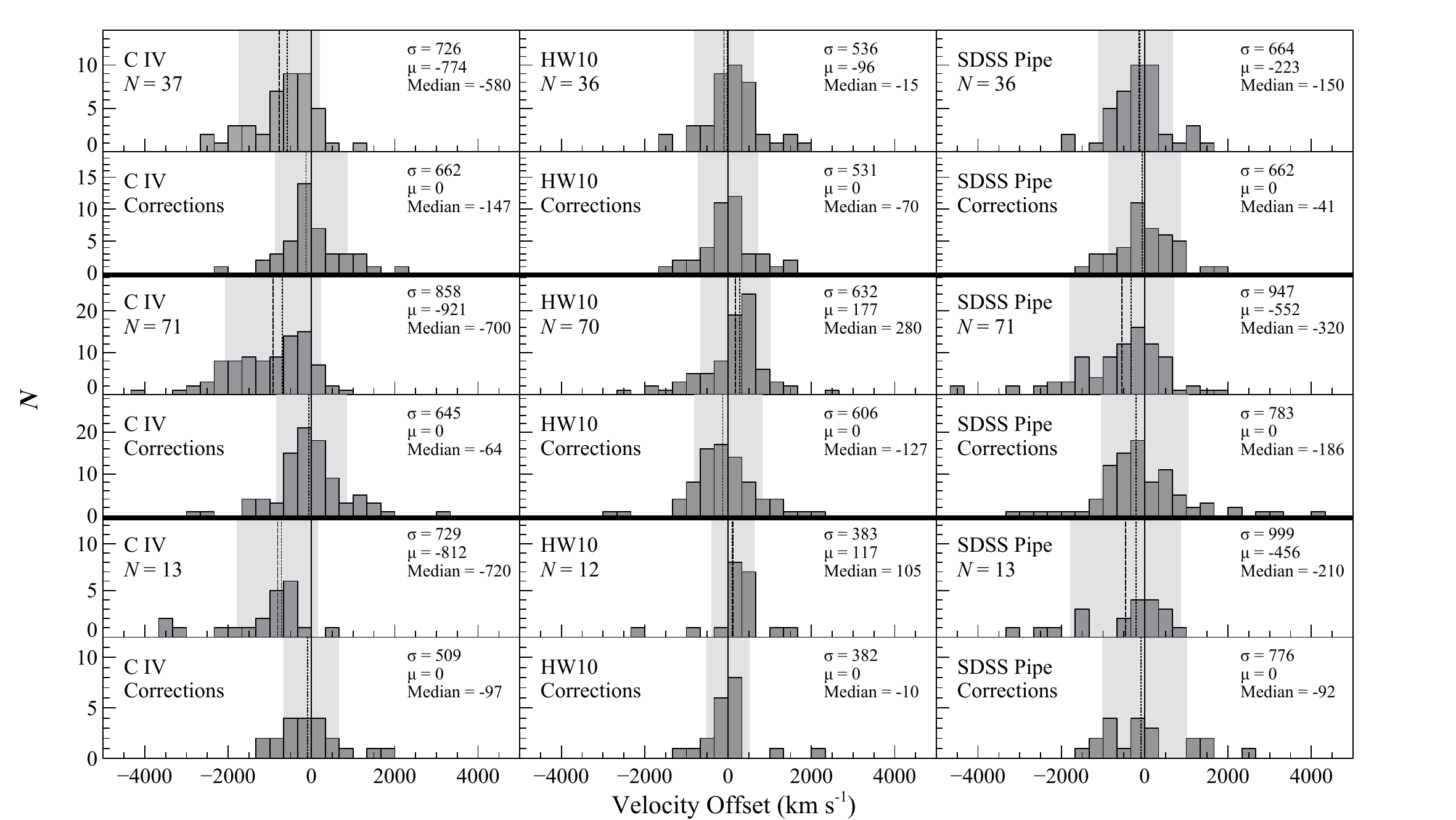}\caption{Same as Figure~\ref{fig:154}, but split into three redshift bins. Top six panels, middle six panels, and bottom six panels correspond to redshift Bin 1, Bin 2, and Bin 3, respectively, as described in the text.}
\label{fig:hists}
\end{figure*}

We perform the regression analysis as described in Section~\ref{sec:data} on each redshift bin separately. The results are presented in Table~\ref{tab:stats}, and shown in Figure~\ref{fig:hists}. The $\sigma$ values of the velocity offsets across all redshift bins are roughly at or below the respective values in the bulk sample with respect to the \cfour\ and HW10 methods (see Figure~\ref{fig:154}). Specifically, while the improvement in $\sigma$ for the bulk sample considering the \cfour -based method is $\sim18\%$, we observe improvements of $\sim9\%$ in Bin 1, $\sim25\%$ in Bin 2, and $\sim30\%$ in Bin 3. Although the statistics in Bin 3 are limited, this trend may follow from the fact that the highest redshift bin tends to have higher luminosity quasars, which results in larger \cfour\ blueshifts (e.g., due to outflows or winds) on average for more distant sources \citep[e.g.,][]{2011AJ....141..167R}. Since our regression analysis relies heavily on the \cfour\ parameter space, it is not unexpected that our corrections to the \cfour -based redshifts would be more important for the more powerful sources found preferentially at higher redshifts.

Concerning the HW10-based method, our corrections produce improvements in standard deviation ranging from $\sim1\%$ to $\sim4\%$, with no apparent trend with redshift. Therefore, it seems that these improvements are not very sensitive to the coverage of the \mgtwo\ line, which is absent from Bin 3. This result may be indicative of the overall robustness of the HW10 method, as found from the entire sample (see Section~\ref{sec:data} and Figure~\ref{fig:154}). Mild improvements, and no significant redshift dependence, are observed for the SDSS Pipeline method, and the overall standard deviations of velocity offset distributions stemming from this method remain high ($\sim 780~\rm{km~s^{-1}}$) in Bins 2 and 3.

\begin{deluxetable*}{cccrrrc}
\tablecaption{Linear Regression Coefficients for Each Redshift Bin \label{tab:stats}}
\tablewidth{0pt}
\tablehead{
\colhead{UV-Based} & \colhead{Redshift} & \colhead{Regression} & \colhead{Value} & \colhead{Error} & \colhead{\textit{t}-Value} & Number of\\
\colhead{Redshift Method} & \colhead{Bin\tablenotemark{a}} & \colhead{Coefficients} & \colhead{} & \colhead{} & \colhead{} & \colhead{Sources}
}
\decimalcolnumbers
\startdata
 & & $\alpha$ & -1103  &  762 & -1.45   \\
 & 1 & $\beta$  & 1022  &  436 & 2.35 & 37\\
 & & $\gamma$ & 37   &  60 & 0.62  \\
 \hline
 & & $\alpha$ & -3382   & 561& -6.02 \\
 \cfour & 2 & $\beta$ & 1357   & 441 & 3.08 & 71\\
 & & $\gamma$ & 204   &  48 & 4.21 \\
 \hline
 & & $\alpha$ & -4802  & 1751& -2.74 \\
 & 3 & $\beta$ & -994  & 1234 & -0.81 & 13\\
 & & $\gamma$ & 391   & 167 & 2.34  \\
 \hline
 \hline
 & & $\alpha$ & -467  &  641 &-0.73  \\
 & 1 & $\beta$  & -29  &  359& -0.08 & 36\\
 & & $\gamma$ & 35 & 50 & 0.71  \\
 \hline
 & & $\alpha$ & -758 &   532& -1.42 \\
 HW10 & 2 & $\beta$ & 717  &  415 & 1.73& 70\\
 & & $\gamma$ & 41 &    46 & 0.89\\
 \hline
 & & $\alpha$ & -27  & 1571 &-0.02 \\
 & 3 & $\beta$ & 85  &  979 & 0.09& 12\\
 & & $\gamma$ & 2 &   148 & 0.01\\
 \hline
 \hline
  & & $\alpha$ & -251 & 780 & -0.32 \\
 & 1 & $\beta$  & 149 & 440 & 0.34 & 36\\
 & & $\gamma$ & 10 & 61 & 0.17 \\
 \hline
 & & $\alpha$ & -2661 & 681 & -3.91 \\
 SDSS Pipe & 2 & $\beta$ & 1825  &  535 & 3.41& 71\\
 & & $\gamma$ & 140   &  59 & 2.38 \\
 \hline
 & & $\alpha$ & 6314  & 2670 & 2.36 \\
 & 3 & $\beta$ & 4486  & 1881 & 2.38& 13\\
 & & $\gamma$ & -643   & 255 &-2.52 \\
 \enddata
 \tablenotetext{a}{Bins 1, 2, and 3 correspond to redshift ranges of \hbox{$1.55 \lesssim z \lesssim 1.65$}, \hbox{$2.10 \lesssim z \lesssim 2.40$}, and \hbox{$3.20 \lesssim z \lesssim 3.50$}.}
\end{deluxetable*}

In general, the greatest limitation in our ability to search for a redshift dependence is the disparity in the number of sources in each bin as is also portrayed by the large uncertainties on the regression coefficients in Table~\ref{tab:stats}. Therefore, we derived the $\Delta v_{\rm corr}$ values in each redshift bin using the coefficients from Table~\ref{tab:coeff1}, and found that the standard deviations on the distributions, when compared to those presented in Figure~\ref{fig:hists}, are roughly consistent across all redshift bins (within $\sim5\%$), which indicates no discernible redshift evolution in our sample. A significantly larger sample size, particularly in Bin 3 ($z \sim 3$), may allow for a more definitive conclusion in this matter. This highest redshift bin is particularly important given the absence of the \mgtwo\ lines from the optical spectrum, and the need to reliably estimate redshifts of more distant sources.

In addition to exploring a possible redshift dependence, we also look to see if our ability to predict a quasar's $z_{\rm sys}$ value depends on source luminosity. We trisect the calibration sample into three $L_{5100}$ ranges with roughly equal widths: $46.08 - 46.41$, $46.42 - 46.74$, and $46.75 - 47.09$, and look for any significant statistical deviations with respect to the entire sample. The results are shown in Figure~\ref{fig:swing}. We find that there appears to be no clear dependence on source luminosity. A possible explanation for this result is that our sample is flux limited, and therefore it is difficult to disentangle the strong redshift-luminosity dependence.

\section{Summary and Conclusions} \label{sec:app}

We present an augmented catalog of spectroscopic properties obtained from NIR observations of a uniform, flux-limited sample of 260 SDSS quasars at \hbox{$1.55 \lesssim z \lesssim 3.50$}. This catalog includes basic spectral properties of rest-frame optical emission lines, chiefly the \mgtwo, \hb, \othree, \fetwo, and \ha ~lines, depending on the availability of the line in the spectrum. These measurements provide an enhancement to the existing GNIRS-DQS database enabling one to more accurately analyze and investigate rest-frame UV-optical spectral properties for high-redshift, high-luminosity quasars in a manner consistent with studies of low-redshift quasars.

We also present prescriptions for correcting UV-based redshifts based on a subset of the GNIRS-DQS sample of 121 sources that are non-BAL, non-RL, have accurate \cfour\ measurements, and have $z_{\rm sys}$ values obtained from prominent \othree\ measurements. We provide measurements of velocity offsets using three different UV-based methods compared to $z_{\rm sys}$ values. This 121-source sample is over twice the size of the calibration sample used in D20, and is both a higher quality and more uniform dataset than M17 and D20.

We attempt to correct for these velocity offsets using a linear regression based on UV continuum luminosity and \cfour\ emission-line properties. Using this approach, we can decrease the standard deviation of the distribution of velocity offsets in our calibration sample by $\sim2\%$ with respect to the best available UV-based redshift method, and by $\sim18\%$ with respect to \cfour -based redshifts. The SDSS Pipeline provides the least precise UV-based redshifts; in particular, the standard deviation of the corrected redshifts is still $\sim20\%$ larger than those achieved for the other two methods. We find that the best way to obtain an accurate and precise $z_{\rm sys}$ value is using the \cfour\ parameter space alone via four basic observables associated with the \cfour\ emission line, and applying the following methodology:

\begin{enumerate}
	\item Measure the observed peak wavelength, EW, and FWHM of \cfour , and the monochromatic luminosity at 1350 \AA~($L_{1350}$).
	\item Calculate an initial redshift measurement, $z_{\rm meas}$, with the observed peak wavelength of \cfour .
	\item Use Equation~\ref{eq:vcorr} and the coefficients in Table~\ref{tab:coeff1} to calculate $\Delta v_{\rm corr}$.
	\item Use Equation~\ref{eq:znew} with the observed $z_{\rm meas}$ and calculated $\Delta v_{\rm corr}$ to obtain a revised, more accurate, and more precise redshift measurement.
\end{enumerate}

Additionally, we explore whether our prescriptions depend on 1) velocity width measurement, of which we determine there is no overt discrepancy based on methodology, 2) source redshift, where we determine that additional data are needed, particularly at the highest redshifts under investigation, in order to obtain more robust results, and 3) source luminosity, where no clear trends are apparent, consistent with the flux-limited nature of our sample.

A primary interest going forward would be bolstering the sample with supplementary observations of quasars, primarily at $z\sim3$, in order to obtain statistically meaningful results on a potential redshift dependence, and further improve UV-based redshift determinations. It will also be interesting to test our prescriptions at the highest accessible redshifts, where considerably larger \cfour\ blueshifts have been observed in sources at $z\gtrsim6$, perhaps due to higher accretion rates \citep[e.g.,][]{2019MNRAS.487.3305M,2020ApJ...905...51S,ha23}. Another avenue of further investigation includes increasing the sample size of quasars with significantly higher spectral resolution, e.g., using Gemini's Spectrograph and Camera for Observations of Rapid Phenomena in the Infrared and Optical \citep[SCORPIO;][]{2020SPIE11447E..74R}, in order to further improve the UV-based redshift corrections by obtaining more accurate line peaks of spectral features. Machine learning can also play an important role as larger data sets will be produced that require redshift correction \textit{en masse}. By utilizing the entire quasar UV spectrum, as opposed to a few key parameters, it will be possible to test if machine learning algorithms can produce more reliable estimates of $z_{\rm sys}$ much more efficiently than our prescriptions allow.

As future projects begin to produce data, we can expect that $\approx 10^{6}$ high-redshift ($z \gtrsim 0.8$) quasars will have redshifts determined through large spectroscopic surveys conducted in the rest-frame UV-optical regime from instruments such as the Dark Energy Spectroscopic Instrument \citep[DESI;][]{2013arXiv1308.0847L,2016arXiv161100036D}, the 4m Multi-Object Spectroscopic Telescope \citep{2012SPIE.8446E..0TD}, and the Subaru Prime Focus Spectrograph \citep[PFS,][]{2016SPIE.9908E..1MT}. For those quasars at \hbox{$1.5 \lesssim z \lesssim 5.0$}, coverage of the \cfour ~emission line will enable crucial redshift corrections, as has been demonstrated in this work. Instruments such as the James Webb Space Telescope \citep[JWST,][]{2006SSRv..123..485G} can provide simultaneous coverage of \cfour, \mgtwo, and \othree, while other facilities can provide other systemic redshift indicators such as [C~{\sc ii}], CO, and Ly$\alpha$ halos \citep[see, e.g.,][]{2018ApJ...854...97D,2019ApJ...887..196F} for $6 \lesssim z \lesssim 9$, allowing for similar investigations of redshift dependencies and corrections for the most distant known quasars.

\begin{acknowledgements}

This work is supported by National Science Foundation grants AST-1815281 (B.~M.~M., C.~D., O.~S.) and AST-1815645 (M.~S.~B., A.~D.~M., J.~N.~M.). W.N.B. acknowledges support from NSF grant AST-2106990. We thank an anonymous referee for thoughtful and valuable comments that helped improve this manuscript. This work was enabled by observations made from the Gemini North telescope, located within the Maunakea Science Reserve and adjacent to the summit of Maunakea. We are grateful for the privilege of observing the Universe from a place that is unique in both its astronomical quality and its cultural significance. This research has made use of the NASA/IPAC Extragalactic Database (NED), which is operated by the Jet Propulsion Laboratory, California Institute of Technology, under contract with the National Aeronautics and Space Administration. We thank Paul Hewett for helpful contributions of redshift data.

\end{acknowledgements}

\appendix

\section{Comparing Different Velocity Widths of the \cfour\ Line} \label{sec:fwhm}

\restartappendixnumbering

\begin{figure*}[h]
\centering
\includegraphics[scale=0.49]{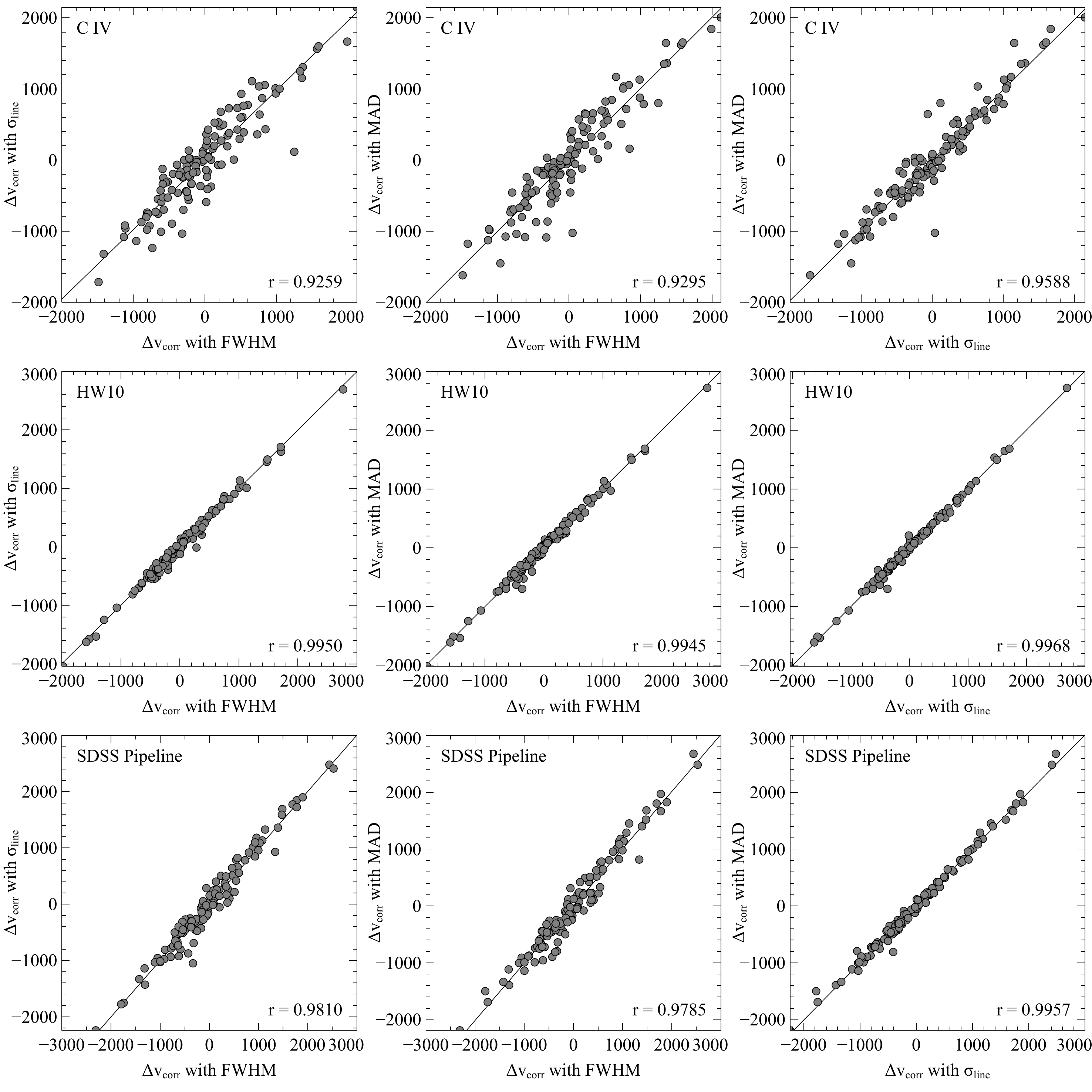}\caption{Comparison of the velocity offsets produced using \cfour ~FWHM, $\sigma_{\rm line}$, and MAD for each UV-based redshift method. Each panel displays the correlation between the corrected velocity offset values produced by our regression analysis when using either FWHM, $\sigma_{\rm line}$, or MAD, along with a corresponding Pearson linear correlation coefficient $r$, where $r \rightarrow 1$ corresponds to a strong correlation. No significant difference exists in this regression analysis between the three different parameters.}
\label{fig:FWLD}
\end{figure*}

In our regression analysis, we have elected to use the FWHM of the \cfour\ line. However, there has been some debate in the literature concerning the overall reliability of using FWHM as the quantification of the velocity width of an emission line \citep[e.g.,][]{2017ApJ...839...93P,2020ApJ...903..112D}. While M17 and D20 used FWHM for their analyses, other methods for measuring velocity widths of emission-line profiles include Line Dispersion ($\sigma_{\rm line}$) and Mean Absolute Deviation (MAD) \citep[e.g.,][]{2016ApJS..224...14D,2020ApJ...903..112D}. We therefore repeated our analysis by replacing FWHM with each of these two velocity width methods, measured from the Gaussian fits presented in Table~\ref{tab:gauss}, and compared the results obtained from all three velocity widths. We find that replacing FWHM with $\sigma_{\rm line}$ or MAD gave no notable improvement in the dispersion on the relevant corrections, as shown in Figure~\ref{fig:FWLD}. We thus have elected to adopt the FWHM parameterization throughout this work.

\end{document}